\documentclass{jfm}

\usepackage{natbib}
\usepackage[normalem]{ulem}
\usepackage[latin1]{inputenc}
\usepackage{afterpage}
\usepackage{graphicx}
\usepackage{longtable}
\usepackage{hyperref}
\usepackage[english]{babel}
\usepackage{color}
\usepackage{dcolumn}
\usepackage{bm}
\usepackage{amsfonts}
\usepackage{amsmath}
\usepackage{amssymb}
\usepackage{amsbsy}
\usepackage{subfigure}
\usepackage{psfrag}
\usepackage{float}
\usepackage{bbm}
\usepackage{soul}

\ifCUPmtlplainloaded \else
  \checkfont{eurm10}
  \iffontfound
    \IfFileExists{upmath.sty}
      {\typeout{^^JFound AMS Euler Roman fonts on the system,
                   using the 'upmath' package.^^J}
       \usepackage{upmath}}
      {\typeout{^^JFound AMS Euler Roman fonts on the system, but you
                   dont seem to have the}%
       \typeout{'upmath' package installed. JFM.cls can take advantage
                 of these fonts,^^Jif you use 'upmath' package.^^J}%
      }
  \else
  \fi
\fi

\ifCUPmtlplainloaded \else
  \checkfont{msam10}
  \iffontfound
    \IfFileExists{amssymb.sty}
      {\typeout{^^JFound AMS Symbol fonts on the system, using the
                'amssymb' package.^^J}%
       \usepackage{amssymb}%
         \let\leq=\leqslant
       \let\ge=\geqslant  \let\geq=\geqslant
      }{}
  \fi
\fi

\ifCUPmtlplainloaded \else
  \IfFileExists{amsbsy.sty}
    {\typeout{^^JFound the 'amsbsy' package on the system, using it.^^J}%
     \usepackage{amsbsy}}
    {\providecommand\boldsymbol[1]{\mbox{\boldmath $##1$}}}
\fi

\newcommand\R{\Re}
\newcommand\I{\Im}
\newcommand\Rey{\mbox{\textit{Re}}}  
\newcommand\Str{\mbox{\textit{St}}}  
\newcommand{\pde}[2]{\frac{\p#1}{\p#2}}

\newsavebox{\astrutbox}
\sbox{\astrutbox}{\rule[-5pt]{0pt}{20pt}}

\newcommand\p{\ensuremath{\partial}}

\title[Global stability analysis of an axisymmetric spinning bullet-like body]
{Global stability analysis of the axisymmetric wake past a spinning bullet-shaped body}

\author[J.I. Jim\'{e}nez-Gonz\'{a}lez, A. Sevilla,  E. Sanmiguel-Rojas and C. Mart\'{\i}nez-Baz\'an]%
{J. I. JIM\'ENEZ-GONZ\'ALEZ$^1$%
\thanks{Email address for correspondence: jignacio@ujaen.es},\ns
A. SEVILLA$^2$\break
E. SANMIGUEL-ROJAS$^3$
\and C. MART\'INEZ-BAZ\'AN$^1$}

\affiliation{$^1$\'Area de Mec\'anica de Fluidos, Departamento de Ingenier\'{\i}a Mec\'anica y Minera. Universidad de Ja\'en, Campus de Las Lagunillas, 23071 Ja\'en, Spain.\\[\affilskip]
$^2$\'Area de Mec\'anica de Fluidos, Departamento de Ingenier\'{\i}a T\'ermica y de Fluidos. Universidad Carlos III de Madrid, 28911 Legan\'es, Spain.\\[\affilskip]
$^3$\'Area de Ingenier\'ia Mec\'anica, Departamento de Mec\'anica. Universidad de C\'ordoba. Campus de Rabanales, 14071 C\'ordoba, Spain.}

\date{?; revised ?; accepted ?. - To be entered by editorial office}

\begin{document}

\maketitle

\begin{abstract}
We analyze the global linear stability of the axisymmetric flow around a spinning bullet-shaped body of length-to-diameter ratio $L/D=2$, as a function of the Reynolds number, $\Rey=w_{\infty}D/\nu$, and of the rotation parameter $\Omega=\omega D/(2 w_{\infty})$, in the ranges $\Rey<450$ and $0\leq\Omega\leq 1$. Here, $w_{\infty}$ and $\omega$ are the free-stream and the body rotation velocities respectively, and $\nu$ is the fluid kinematic viscosity. The two-dimensional eigenvalue problem is solved numerically to find the spectrum of complex eigenvalues and their associated eigenfunctions, allowing us to explain the different bifurcations from the axisymmetric state observed in previous numerical studies. Our results reveal that, for the parameter ranges investigated herein, three global eigenmodes, denoted Low-Frequency (LF), Medium-Frequency (MF) and High-Frequency (HF) modes, become unstable in different regions of the $\Rey-\Omega$ parameter plane. We provide precise computations of the corresponding neutral curves, that divide the $\Rey-\Omega$ plane into four different regions: the stable axisymmetric flow prevails for small enough values of $\Rey$ and $\Omega$, while three different frozen states, where the wake structures co-rotate with the body at different angular velocities, take place as a consequence of the destabilization of the LF, MF and HF modes. Several direct numerical simulations of the nonlinear state associated to the MF mode, identified here for the first time, are also reported to complement the linear stability results. Finally, we point out the important fact that, since the axisymmetric base flow is $SO(2)$-symmetric, the theory of equivariant bifurcations implies that the weakly non-linear regimes that emerge close to criticality must necessarily take the form of \emph{rotating-wave states}. These states, previously referred to as \emph{frozen wakes} in the literature, are thus shown to result from the base-flow symmetry.
\end{abstract}

\section{Introduction}\label{Int}

Unstable modes in wakes behind axisymmetric bluff bodies aligned with the outer flow have been the aim of several studies in the past, and it is now well known that the sequence of transitions towards unsteadiness in the laminar regime is governed by the destabilization of helical modes with azimuthal wavenumber $|m|=1$~\citep[see][among others]{Achenbach74,Monkewitz88b,Natarajan93,Tomboulides00,Pier08,Sanmiguel09}. Regarding wakes behind rotating bluff bodies, \cite{Pier13} performed a linear stability analysis showing that the helical mode is responsible for the instability of the wake of a spinning sphere. Moreover, he showed that this mode leads to a co-rotating \emph{frozen} flow~\citep{Kim02}, where the vortices rotate without change in shape nor in intensity, resembling results previously reported for other swirling flows. For instance, it is well known that spiral vortex breakdown occurs as a consequence of the destabilization of a helical global mode, characterized by a co-rotating wave~\citep{Gallaire06,Oberleithner11}. Similarly,~\cite{Khorrami91} found through a temporal stability analysis that the Batchelor vortex is unstable to a long-wave co-rotating helical mode for moderate values of the Reynolds number and the swirl parameter. However, the stability of the flow past spinning bodies still needs to be thoroughly addressed.

Recent numerical simulations have revealed important differences between the stability properties of the wake of a spinning bullet-shaped body~\citep[][]{Jimenez13}, and that of a rotating sphere~\citep{Kim02,Pier13}. Indeed, while the axisymmetric state is stabilized by a moderate amount of rotation for the bullet-shaped body, it is destabilized in the case of the sphere. Such differences are apparently related to the stabilizing effect associated with the body base, allowing recovery of the axisymmetry of the wake under certain conditions. However, the behaviour of this stabilizing effect for bullet-like bodies at high rotation parameters has not been explored, since the determination of the bifurcations by means of three-dimensional numerical simulations is an extremely time-consuming task in marginally stable cases. Thus, in the present work we use a global linear analysis to explore the stability limits of the axisymmetric steady state for moderate and high values of the rotation parameter and low Reynolds numbers, while also providing valuable information about the physics of the problem at a considerably smaller computational cost.

As already mentioned, the application of rotation to a bullet-shaped body allows retrieval of the axisymmetry within certain ranges of the rotation velocity and Reynolds number, suggesting its use as a passive control method. Therefore, establishing the region of the $(\Omega,\Rey)$-plane where the axisymmetric flow is stable has great interest from a practical point of view, since its boundaries provide the critical values for which the flow past the spinning body is completely stabilized. To this end, the global stability analysis technique seems an appropriate tool, since it has been proved to accurately predict the bifurcations from the axisymmetric state in non-rotating bullet-shaped bodies~\citep{Bohorquez11}. In addition, the modal solutions of the Navier-Stokes equations linearized around the basic axisymmetric state constitute, together with the corresponding adjoint problem, the first step towards a deeper understanding of the sensitivity and receptivity of the flow, aimed at finding efficient
control strategies~\citep{Chomaz05,Sipp10,Tchoufag13}. Therefore, the present work could also be used in the future as a starting point for a more complete study, where the control of wakes behind spinning bodies is achieved by slightly perturbing the base flow according to its sensitivity characteristics.

This paper begins with the problem formulation and details of the numerical techniques, which are presented in \S~\ref{Formulation}. Results and discussions about the leading global modes and their corresponding stability boundaries within the ranges $0\leq\Omega\leq1$ and $\Rey<450$, are included in \S~\ref{SectRESULTS}. More specifically, in \S~\ref{SectRESULTSa}, we show the results for $\Omega\leq 0.4$, which have been used to validate the stability code for spinning bodies, comparing its results with previous three-dimensional numerical simulations. In \S~\ref{SectRESULTSb} we include results for $0.4<\Omega\leq 1$, obtained by means of the stability analysis, as well as several three-dimensional numerical simulations performed to complement our previous work and to identify new nonlinear bifurcated regimes, allowing us to define a complete bifurcation diagram in the ($\Omega,\Rey$)-parametric plane. Finally, a physical interpretation of the results obtained is given in \S~\ref{SectRESULTSc}, while main
conclusions are outlined in \S~\ref{SectCONCLUSIONS}.

\section{Problem formulation}\label{Formulation}

\begin{figure}
\begin{center}
\includegraphics[width=14 cm]{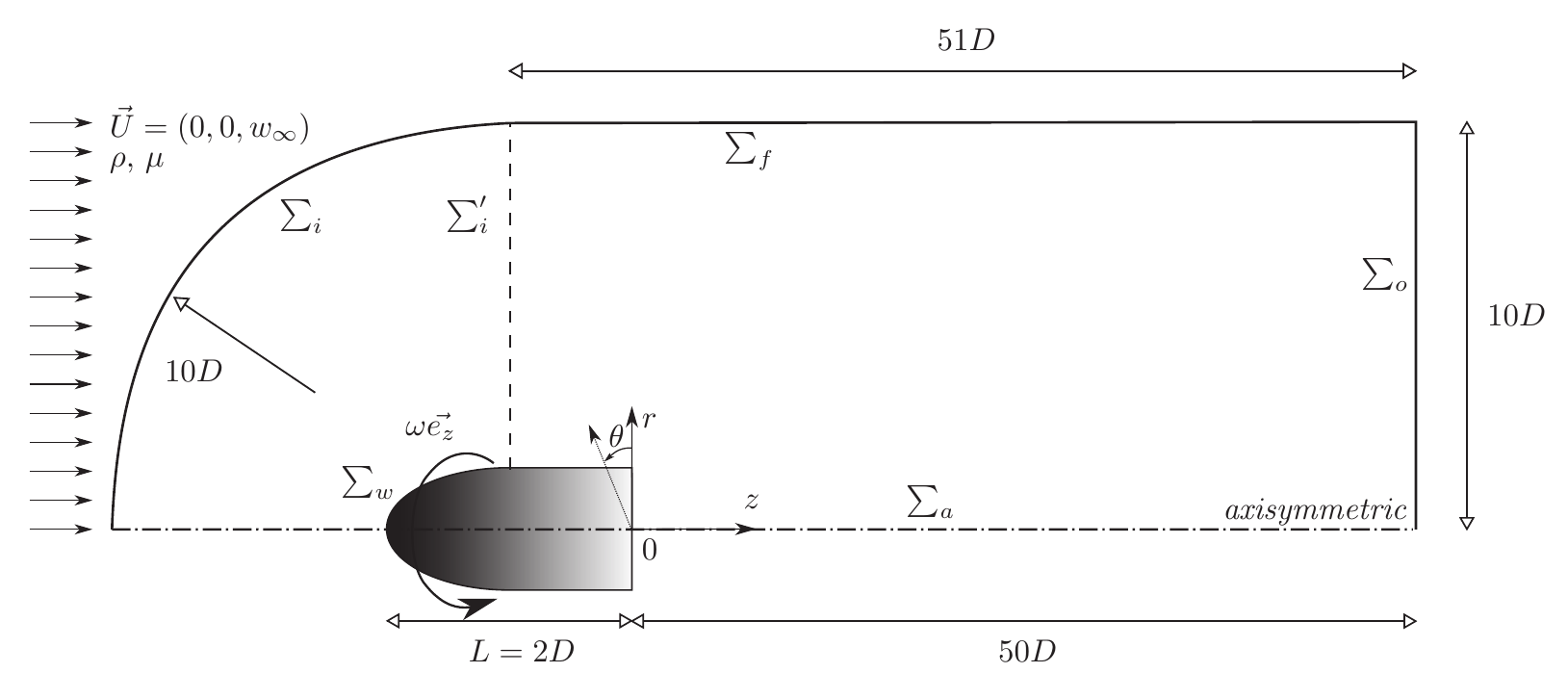}
\caption{Scheme of the problem and axisymmetric computational domain. The vertical dashed line represents the inlet for the mesh used in the global stability analysis.}
\label{scheme}
\end{center}
\end{figure}

The flow configuration, sketched in figure~\ref{scheme}, consists of a uniform stream of velocity $w_{\infty}$, density $\rho$ and viscosity $\mu$, around a bullet-like body of length-to-diameter ratio $\ell=L/D=2$, that rotates about its axis with a constant angular velocity $\omega$. The equations governing the flow are the three-dimensional, incompressible Navier-Stokes equations,
\begin{gather}
\nabla\cdot\mathbf{u}=0, \label{NV1}\\
\pde{\mathbf{u}}{t}+\mathbf{u}\cdot\nabla\mathbf{u}+\nabla
p-\frac{1}{\Rey}\nabla^2\mathbf{u}=0\label{NV2},
\end{gather}
being $\mathbf{u}=(u,v,w)$ and $p$ the dimensionless velocity and pressure fields expressed in cylindrical coordinates $(r,\theta,z)$. Here the variables have been made dimensionless using $D$, $w_{\infty}$, $D/w_{\infty}$ and $\rho w_{\infty}^2$ as scales of length, velocity, time and pressure, respectively. Thus, two non-dimensional parameters characterize the problem, namely the Reynolds number, $\Rey=\rho w_{\infty}D/\mu$, and the rotation parameter, $\Omega=\omega D/(2w_{\infty})$.

Although, in the present investigation, we have solved numerically the full unsteady, three-dimensional problem given by equations~\eqref{NV1}-\eqref{NV2} in several representative cases, we have mainly focused on the global linear stability problem, formulated by means of the following standard decomposition,
\begin{equation}
[\mathbf{u}(\mathbf{x},t),p(\mathbf{x},t)]=\left[\mathbf{U}(\mathbf{x})+
\mathbf{u'}(\mathbf{x},t),P(\mathbf{x})+p'(\mathbf{x},t)\right].\label{superp}
\end{equation}
The basic velocity, $\mathbf{U}(\mathbf{x})=\left[U(r,z),V(r,z),W(r,z)\right]$, and pressure, $P(r,z)$, fields are the solution of the steady, axisymmetric version of equations~\eqref{NV1} and~\eqref{NV2},
\begin{gather}
\nabla\cdot\mathbf{U}=0,\label{BF1}\\
\mathbf{U}\cdot\nabla\mathbf{U}+\nabla
P-\frac{1}{\Rey}\nabla^2\mathbf{U}=0,\label{BF2}
\end{gather}
together with appropriate boundary conditions discussed below, while the small-amplitude disturbances $[\mathbf{u'}(\mathbf{x},t),p'(\mathbf{x},t)]$ satisfy the unsteady Navier-Stokes equations linearized around the base state, namely
\begin{gather}
\nabla\cdot\mathbf{u'}=0, \label{PF1}\\
\frac{\partial\mathbf{u'}}{\partial
t}+\mathbf{U}\cdot\nabla\mathbf{u'}+\mathbf{u'}\cdot\nabla\mathbf{U}+\nabla
p'-\frac{1}{\Rey}\nabla^2\mathbf{u'}=0,\label{system}
\end{gather}
that in compact form, read
\begin{equation}
-\pde{}{t}\mathbf{q'}=\mathcal{L}(\mathbf{U})\mathbf{q'}.\label{IVP}
\end{equation}
Equation~\eqref{IVP}, subjected to the corresponding set of boundary conditions, constitutes an initial value problem, where $\mathbf{q'}=(\mathbf{u'},p')$, and $\mathcal{L}(\mathbf{U})$ denotes the linear stability operator. In the present work we are only concerned with the modal problem, i.e. with the eigenmodes associated to equation~\eqref{IVP}. Furthermore, since the base flow is steady and axisymmetric, the resulting eigenvalue problem is two-dimensional, and is formulated by means of the ansatz
\begin{equation}
[\mathbf{u'},p']=\left[\mathbf{\hat{u}}(r,z),\hat{p}(r,z)\right]e^{\sigma t+i m\theta},
\label{exponential}
\end{equation}
where $\sigma_r=\Re(\sigma)$ is the growth rate of each mode, $\sigma_i=\Im(\sigma)$ its angular frequency, $m \in \mathbb{Z}$ the azimuthal wavenumber, and $(\mathbf{\hat{u}},\hat{p})$ the global eigenfunctions. Introducing equation~\eqref{exponential} into~\eqref{IVP} yields the following generalized eigenvalue problem (EVP),
\begin{equation}
\mathcal{A}\mathbf{\hat{q}}=\sigma\mathcal{B}\mathbf{\hat{q}},\label{eigenproblem}
\end{equation}
where $\mathbf{\hat{q}}=(\mathbf{\hat{u}},\hat{p})^T=(\hat{u},\hat{v},\hat{w},\hat{p})^T$
is the eigenfunction vector and, $\mathcal{A}$ and $\mathcal{B}$ are the following linear
operators
\begin{gather}\label{matrixA}
\noindent \hspace{-13 cm} \mathcal{A}= \\ \nonumber
\noindent \makebox[13.1cm]{\small
\(\displaystyle \begin{pmatrix}\pde{}{r}+\frac{1}{r} & \frac{im}{r}
&\pde{}{z} & 0\\
\noalign{\vskip4pt}
\mathbf{U}\cdot\nabla+\pde{U}{r}-\frac{1}{\Rey}\left(\nabla ^2 -
\frac{1}{r^2}\right) &
-\frac{2V}{r}+\frac{2}{\Rey}\frac{im}{r^2} & \pde{U}{z} & \pde{}{r}\\
\noalign{\vskip4pt}
\pde{V}{r}+\frac{V}{r}-\frac{2}{\Rey}\frac{im}{r^2} &
\mathbf{U}\cdot\nabla+\frac{U}{r}-\frac{1}{\Rey}\left(\nabla ^2 -
\frac{1}{r^2}\right) & \pde{V}{z} & \frac{im}{r}\\
\noalign{\vskip4pt} \pde{W}{r} & 0 &
\mathbf{U}\cdot\nabla+\pde{W}{z}-\frac{1}{\Rey}\nabla ^2 & \pde{}{z}
\end{pmatrix},\)}\nonumber
\end{gather}

\begin{gather}
\mathcal{B}=-\begin{pmatrix}
O & O & O & O\\
O & I & O & O\\
O & O & I & O\\
O & O & O & I
\end{pmatrix},
\label{matrixB}
\end{gather}\\
where $O$ and $I$ are the null and identity matrices, respectively. By numerically solving equation~\eqref{eigenproblem} the leading global modes $\mathbf{\hat{q}}$ of the axisymmetric flow around the bullet-shaped body can be identified, together with the stability limits of the axisymmetric steady state and the nature of the bifurcations that take place in the ranges of $\Rey$ and $\Omega$ under study. Numerical details on the base flow computation and the discretization and solution of the generalized EVP can be found in Appendix~\ref{sec:numerics}.

\section{Results and discussion}\label{SectRESULTS}

In this section we present the stability results obtained for a range of rotation parameters $0\leq\Omega\leq 1$, which are varied in increments of $\Delta\Omega=0.05$ (or $0.01$ within some distinguished regions). To properly capture the transitions from the axisymmetric wake towards other unstable states, different ranges of Reynolds numbers have been covered for each value of $\Omega$, with small increments of $\Delta \Rey=10$ to accurately track the bifurcations that take place for increasing values of the rotation parameter. Additionally, to validate the results of the stability analysis obtained for $\Omega>0.4$, direct numerical simulations were also performed in the parametric ranges $0.4<\Omega\leq 0.6$ and $\Rey\leq420$. Finally, a discussion about the structure of the eigenfunctions and the nature of the unstable modes is included.
\begin{figure}
\begin{center}
\includegraphics[width=12cm]{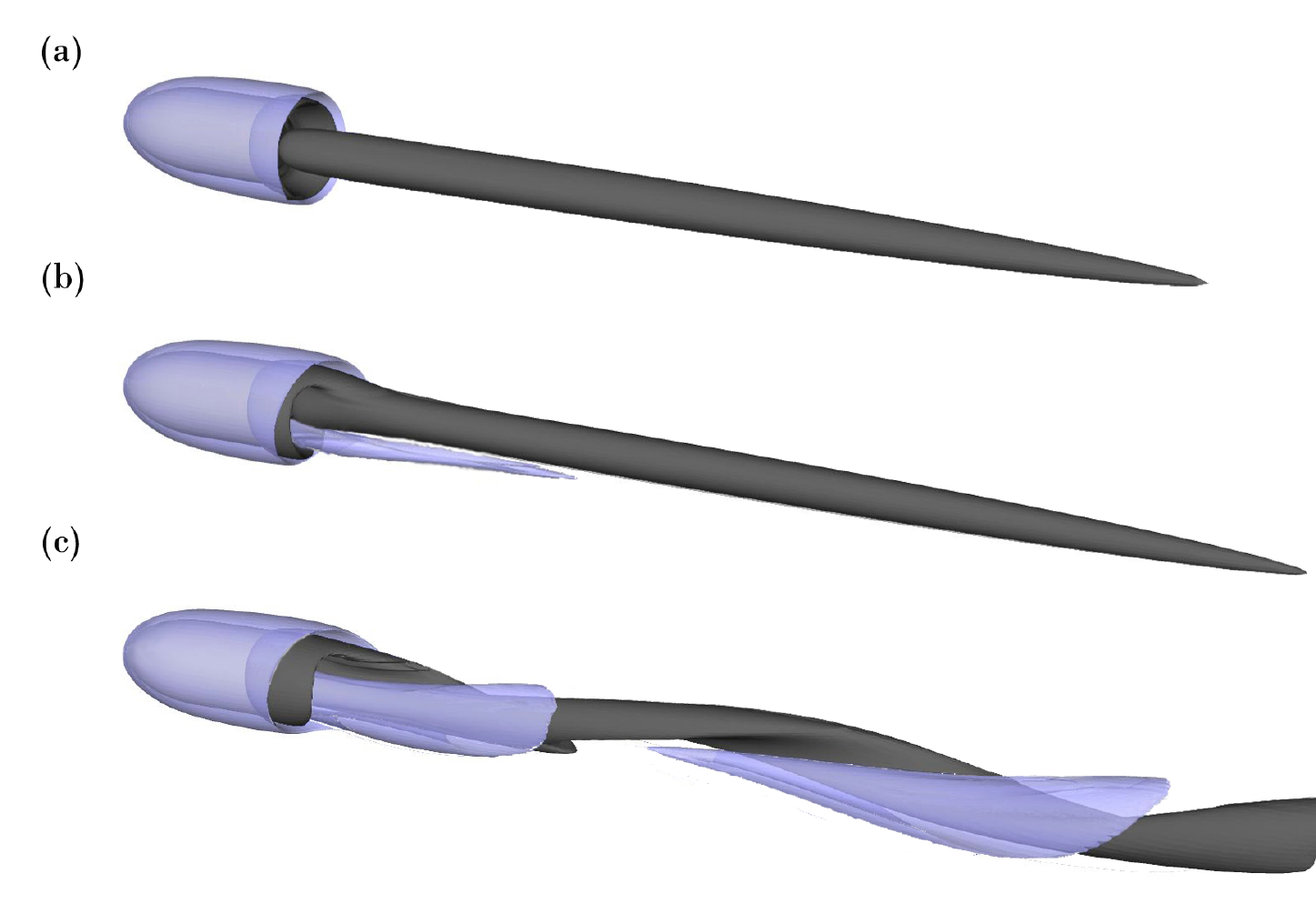}
\caption{Contours of constant streamwise vorticity, $\varpi_z=0.15$
(dark-coloured contour) and $\varpi_z=-0.15$ (light-coloured
contour), for a body of aspect ratio $\ell=2$ rotating at
$\Omega=0.15$, for (a) $\Rey=330$, (b) $\Rey=370$ and (c)
$\Rey=430$.} \label{0p1}
\end{center}
\end{figure}

\subsection{Nonlinear regimes for $\Omega\leq 0.4$}\label{SectRESULTS0}
Let us first briefly summarize the phenomenology of the different flow bifurcations that take place in the wake of bullet-like bodies at increasing Reynolds numbers. The axisymmetric steady wake behind a bullet-shaped body of aspect ratio $\ell=2$ without rotation undergoes two subsequent bifurcations at increasing Reynolds numbers~\citep{Bohorquez11}. The first regular bifurcation leads to a planar-symmetric \emph{Steady State} (SS) mode at $\Rey_{c1}\simeq 319$~\citep{Fabre08}, featuring two counter-rotating vortices aligned with the flow. The second Hopf bifurcation takes place at $\Rey_{c2}\simeq 412$, where the flow evolves towards an oscillatory vortex shedding regime that retains the planar symmetry, being denoted \emph{Reflectional Symmetry Preserving} (RSP) mode~\citep{Fabre08}. Body rotation modifies the values of $\Rey_{c1}$ and $\Rey_{c2}$, which become functions of $\Omega$, as well as the spatial structure of the flow, as can be seen in figure~\ref{0p1} for $\Omega=0.1$~\citep[see][]{Jimenez13}.
In fact, at $\Rey_{c1}(\Omega)>\Rey_{c1}(\Omega=0)$, a modified unsteady SS mode arises (figure~\ref{0p1}b), where the two counter-rotating vortices are no longer symmetric, since the thread of positive axial vorticity, $\varpi_z$, becomes more intense due to the rotation of the body base, while the negative one is weakened. The main feature of this unsteady mode is that the vortical structures rotate with an angular velocity (which is an increasing function of $\Omega$) different from that of the body, without variation in strength or shape, being thus denoted as \emph{Frozen State}~\citep[similarly to the regime described for the sphere by][]{Kim02}. When the second transition takes place at $\Rey_{c2}(\Omega)$, an oscillatory mode develops but, in this case, it is spirally deformed by the body rotation (figure~\ref{0p1}c), being thus denoted \emph{Spiral Unsteady State}. The values of both $\Rey_{c1}$ and $\Rey_{c2}$ are highly affected by the spin of the body, that stabilizes the frozen mode (SS mode at
$\Omega=0$), such that the value of $\Rey_{c1}$ is an increasing function of $\Omega$. In contrast, spin slightly destabilizes the unsteady spiral mode, such that both bifurcations collapse onto a single one at $\Omega\approx 0.225$.
\begin{figure}
\begin{center}
\includegraphics[width=12cm]{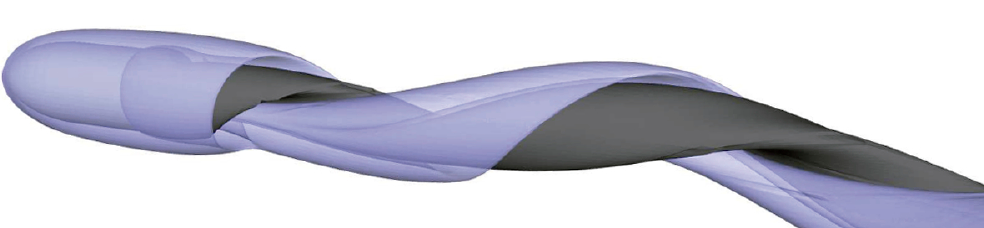}
\caption{Contours of constant streamwise vorticity, $\varpi_z=0.15$
(dark-coloured contour) and $\varpi_z=-0.15$ (light-coloured
contour), for a body of aspect ratio $\ell=2$ rotating at
$\Omega=0.2$ for $\Rey=430$, showing the frozen spiral
mode.\label{spiral_frozen}}
\end{center}
\end{figure}

At $\Omega=0.2$ and increasing $\Rey$, the flow undergoes a third bifurcation at $\Rey=\Rey_{c3}>\Rey_{c2}$ towards a different \emph{Frozen Spiral regime}, depicted in figure~\ref{spiral_frozen}, and characterized by the rotation of the flow structures in the wake without change in shape or intensity, but with an angular frequency much larger than that associated with the frozen state arising at $\Rey_{c1}$ (Figure \ref{0p1}b). Interestingly, this frozen spiral mode represents the only unstable state existing for $\Omega\geq0.225$, in such a way that the wake undergoes a Hopf bifurcation from the axisymmetric state to the frozen spiral regime at $\Rey_{c3}$, whose value seems
to be independent of $\Omega$.

\subsection{Stability characteristics for $\Omega\leq 0.4$}\label{SectRESULTSa}

In the case of a rotating body studied in the present work, the only unstable global modes that have been found, at least within the ranges of $\Rey$ and $\Omega$ investigated, have $m=\pm 1$, while those with $m= 0$ and $m=\pm 2$ are always stable. It is important to point out that in all the calculations reported in the present paper we have set $m=-1$ without loss of generality, since the eigenvalue problem defined in equation~\eqref{eigenproblem} is invariant under the transformation $(m,\sigma,\mathbf{\hat{q}})\to(-m,\sigma^*,\mathbf{\hat{q}^*})$, where the asterisk denotes the complex conjugate.
\begin{figure}
\begin{center}
\includegraphics[width=13cm]{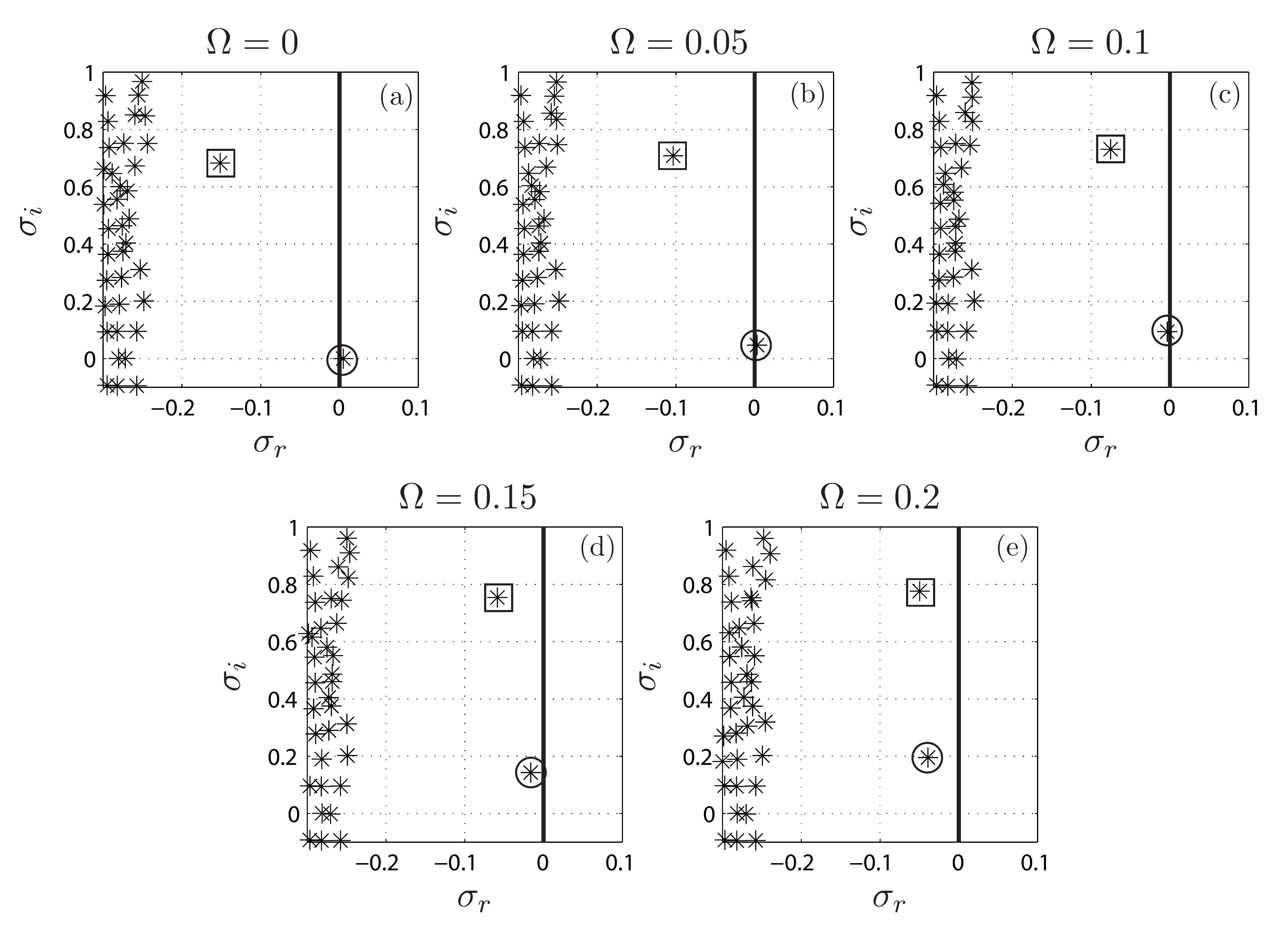}
\caption{Eigenvalue spectra for $\Rey=330$, $m=-1$ and: (a) $\Omega=0$, (b) $\Omega=0.05$,
(c) $\Omega=0.1$, (d) $\Omega=0.15$ and (e) $\Omega=0.2$. The circled mode stands for the
LF mode (SS mode at $\Omega=0$), while the squared one represents the HF mode.}
\label{Ome_leq_0p2}
\end{center}
\end{figure}

Our results show that rotation has a stabilizing effect at low rotation numbers. As an example, the effect of rotation on the stability of the axisymmetric state for $\Rey=330$ and $m=-1$ is illustrated in figure~\ref{Ome_leq_0p2}, where the global eigenvalue spectrum is plotted for several values of $\Omega$ in the range $0\leq \Omega\leq 0.2$. In the case without rotation, shown in figure~\ref{Ome_leq_0p2}(a), there is only one unstable eigenvalue (circled), whose imaginary part is zero and is responsible for the SS mode. The first distinguished stable eigenmode appearing in figure~\ref{Ome_leq_0p2}(a) (squared) is oscillatory, being related with the RSP mode that appears at higher values of the Reynolds number. In the presence of body rotation, figures~\ref{Ome_leq_0p2}(b)-(e) reveal that the leading eigenmode is no longer steady, as happens in the case without rotation, being substituted by an eigenmode with a non-zero imaginary part that will hereafter be referred to as \emph{Low Frequency mode} or LF
mode. The instability of the LF mode (arising from a Hopf bifurcation) is responsible for the bifurcation towards the \emph{Frozen State} described above (see figure~\ref{0p1}b). Moreover, as expected from the characteristics of the frozen state described in \S \ref{SectRESULTS0}, the angular frequency of the LF mode, $\sigma_i$, increases with $\Omega$, and it is stabilized by rotation, since increasing $\Omega$ decreases the real part of the corresponding eigenvalue, $\sigma_r$. For the particular value of $\Rey=330$, the LF mode becomes neutrally stable at a critical value of $\Omega\simeq 0.1$ (figure~\ref{Ome_leq_0p2}c). Thus, the linear stability analysis predicts that the axisymmetric state is stable for $\Rey=330$ and $\Omega\gtrsim 0.1$. On the other hand, the oscillatory mode related with the RSP state in the case without rotation (squared eigenvalue in figure~\ref{Ome_leq_0p2}) is destabilized by the spin together with an increase in its frequency, since the corresponding values of $\sigma_r$ and
$\sigma_i$ increase with $\Omega$. Considering that reflectional symmetry is not preserved when the body spins, it seems reasonable to denote this eigenmode as the \emph{High Frequency mode} or HF mode.
\begin{figure}
\begin{center}
\includegraphics[width=14cm]{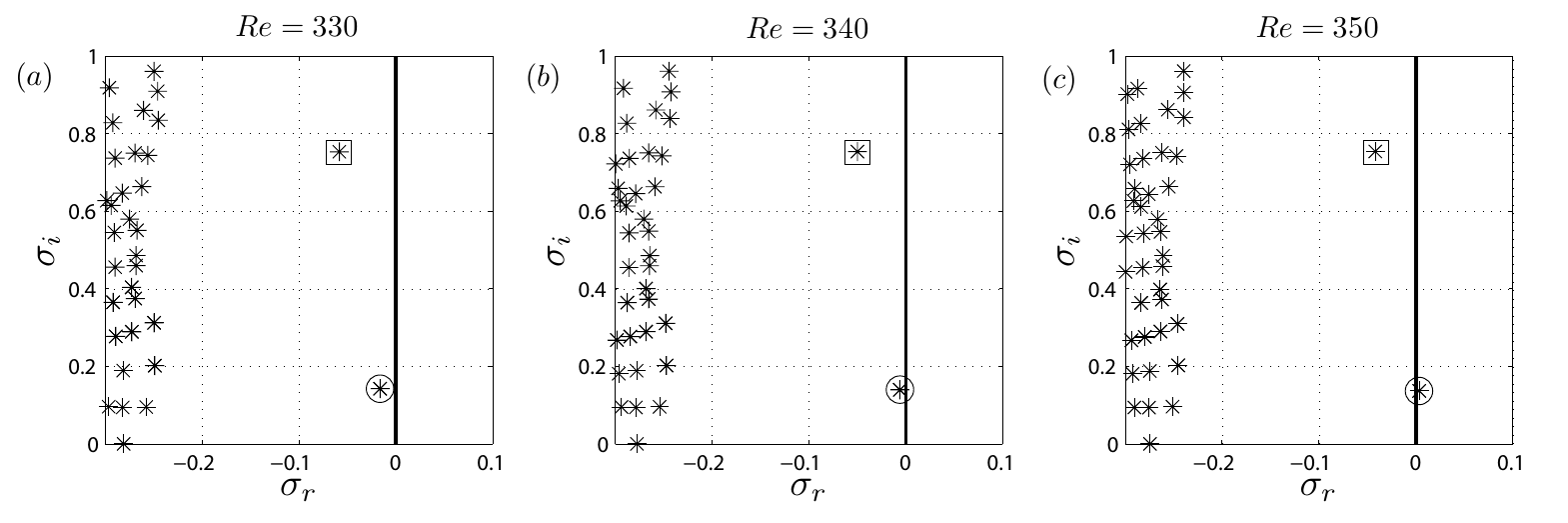}
\caption{Eigenvalue spectra for $\Omega=0.15$, $m=-1$ and:
(a) $\Rey=330$, (b) $\Rey=340$ and (c) $\Rey=350$.} \label{Ome0p15}
\end{center}
\end{figure}
\begin{table}
\centering
\begin{tabular}{c c c c c c c}
    $\Omega$  &  $\Rey_{c1}^{GLS}$  &  $\Str^{GLS}$  &  $\Rey_{c1}^{DNS}$  &  $\Str^{DNS}$ &
    $\epsilon_{\Rey_{c1}}(\%)$ &  $\epsilon_{\Str}(\%)$\\ \\
     $0$  &  325.21 &  0  &  319 &  0 & 1.9467 & 0 \\
     $0.05$  &  326.99  &  0.00743  &  323.5   &  0.00772  & 1.0788 & 3.7565 \\
     $0.10$  &  333.56  &  0.01466  &  332.3  &  0.01532 & 0.3792 & 4.3081 \\
     $0.15$  &  346.49  &  0.02182  &  347.5  &  0.02298 & 0.2906 & 5.0479 \\
     $0.20$  &  370.20 &  0.02802  &  375.4  &  0.02872 & 1.3852 & 2.4373 \\
\end{tabular}
\caption{The values of $\Rey_{c1}$ and $St$ predicted by the global stability analysis
(GLS) for different values of $\Omega$ ($St$ obtained at $\Rey=330,340,350$ and $380$
for $\Omega=0.05,0.1,0.15$ and $0.2$, respectively), compared with those obtained numerically by~\cite{Jimenez13} (DNS). The relative errors,
$\epsilon_{f}(\%)=|f_{GLS}-f_{DNS}|/f_{DNS}\times100$, being $f$ either $\Rey_{c1}$ or
$\Str$, are also shown.}\label{Rec1}
\end{table}

\begin{figure}
\begin{center}
\includegraphics[width=13.5cm]{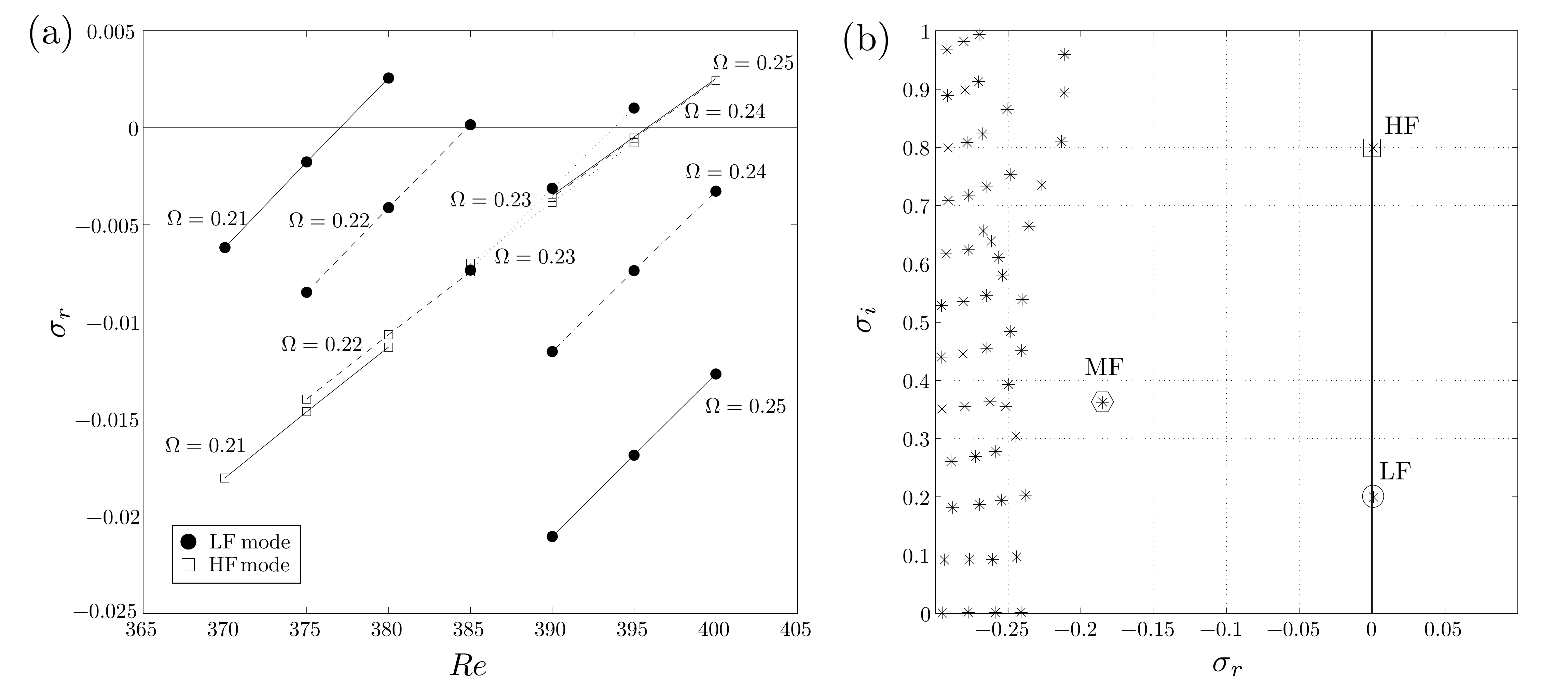}
\caption{(a) Growth rate, $\sigma_r$, as a function of $\Rey$ for the two leading eigenvalues, namely the LF mode ($\bullet$) and the HF mode ($\square$), for $0.21\leq\Omega\leq 0.25$ and $m=-1$ (binding lines represent linear fits); and (b) eigenvalue spectrum for $\Omega=0.232$, $\Rey=395$ and $m=-1$.\label{0p2_0p25}}
\end{center}
\end{figure}

Figure~\ref{Ome0p15} reveals that, for a constant value of $\Omega=0.15$, the flow is destabilized as $\Rey$ increases, since the real part of the leading eigenmode (LF mode) becomes larger and eventually crosses the imaginary axis at a critical value of the Reynolds number $\Rey_{c1}$ (likewise, the HF mode becomes less stable as $\Rey$ increases, as figure~\ref{Ome0p15} shows). To compute the value of $\Rey_{c1}$ we take advantage of the fact that, close to the transition, the growth rate $\sigma_r$ of the leading eigenvalue increases linearly with $\Rey$, as pointed out in~\cite{Bouchet06} (see also figure~\ref{0p2_0p25}a). Thus, we determine $\Rey_{c1}$ by linearly interpolating the values of $\sigma_r$ obtained for two different values of $\Rey$ close to the marginal conditions. In the particular case with $\Omega=0.15$ (figure~\ref{Ome0p15}), the critical Reynolds number is $\Rey_{c1}(\Omega=0.15)\simeq 346.5>\Rey_{c1}(\Omega=0)\simeq 325.2$, so that rotation stabilizes the first bifurcation. From the stability results it is also deduced that the frequency of the LF mode, $\sigma_i$, depends weakly on $\Rey$ for a fixed value of $\Omega$. Moreover, for $\Rey>\Rey_{c1}$, the values of the Strouhal number predicted by the stability analysis, namely $St=\sigma_i/2\pi$, are in good quantitative agreement with those extracted from the numerical simulations close to criticality~\citep[see][and Table~\ref{Rec1}]{Jimenez13}. Consequently, our linear stability analysis is able to properly capture the main features of the first bifurcation from the axisymmetric flow to the frozen state. Indeed, for comparison, table~\ref{Rec1} shows the values of $\Rey_{c1}$ obtained from the numerical simulations, $\Rey_{c1}^{DNS}$, and from the stability analysis, $\Rey_{c1}^{GLS}$, for different values of the rotation parameter $\Omega$. Also shown in table~\ref{Rec1} are the values of $St$ under slightly supercritical values of the Reynolds number corresponding to each value of $\Omega$, as well as the relative errors in both predictions.

The picture described above applies to a range of values of the rotation parameter $0<\Omega\lesssim 0.2$, for which the LF mode is the only relevant one to understand the bifurcation scenario at increasing $\Rey$. However, as pointed out before (see figure~\ref{Ome_leq_0p2}), at $\Omega=0.2$ the growth rates of the LF and HF modes are similar for a slightly subcritical Reynolds number $\Rey=330<\Rey_{c1}$. Thus, it is expected that the HF mode plays a relevant role at higher values of $\Omega$. In fact, as described in \S~\ref{SectRESULTS0}, for $\Omega=0.2$ and increasing $\Rey$, the flow undergoes a bifurcation at $\Rey=\Rey_{c3}$ towards a \emph{Frozen Spiral regime}. Hence, a natural explanation for the onset to this regime, is that the HF mode becomes destabilized at a smaller value of $\Rey$ than the LF one for sufficiently large values of $\Omega$ (in particular, for $\Omega \ge 0.225$ according to the numerical results, and for $\Omega \ge 0.232$ according to the global stability analysis).
Indeed, figure~\ref{0p2_0p25}(a) shows the dependence of $\sigma_r$ with $\Rey$ for both the LF and HF modes, calculated for several values of the rotation parameter in the range $0.2\leq\Omega\leq 0.25$. The results of figure~\ref{0p2_0p25}(a) reveal that, for a certain value of the rotation parameter $\Omega_{c1}$ within the range $0.23<\Omega_{c1}<0.24$, the LF and HF modes become unstable at the \emph{same} critical value of the Reynolds number, i.e. $\Rey_{c1}(\Omega_{c1})=\Rey_{c3}(\Omega_{c1})$, where we have denoted $\Rey_{c3}$ as the critical Reynolds number at which the HF mode is destabilized, according to the corresponding definition in~\S~\ref{SectRESULTS0}. Moreover, it is also deduced from figure~\ref{0p2_0p25}(a) that $\Rey_{c1}<\Rey_{c3}$ (resp. $\Rey_{c1}>\Rey_{c3}$) for $\Omega<\Omega_{c1}$ (resp. $\Omega>\Omega_{c1}$). Therefore, the first bifurcation taking place in the flow at increasing Reynolds numbers crucially depends on the value of $\Omega$, such that the LF and HF modes are first
destabilized for $\Omega<\Omega_{c1}$ and $\Omega>\Omega_{c1}$, respectively. The analysis of the eigenvalues provides the values $\Omega_{c1}\simeq 0.232$ and $\Rey_{c1,3}\simeq 395$, as figure~\ref{0p2_0p25}(b) depicts. Another noteworthy feature of the spectrum displayed in figure~\ref{0p2_0p25}(b) is the existence of a third distinguished eigenvalue in the stable half-plane ($\sigma_r<0$), whose frequency falls between those of the LF and HF modes. This new mode will be denoted as \emph{Medium Frequency mode} or MF mode, and will be shown below to become dominant at large
enough values of $\Omega$.
\begin{figure}
\begin{center}
\includegraphics[width=10cm]{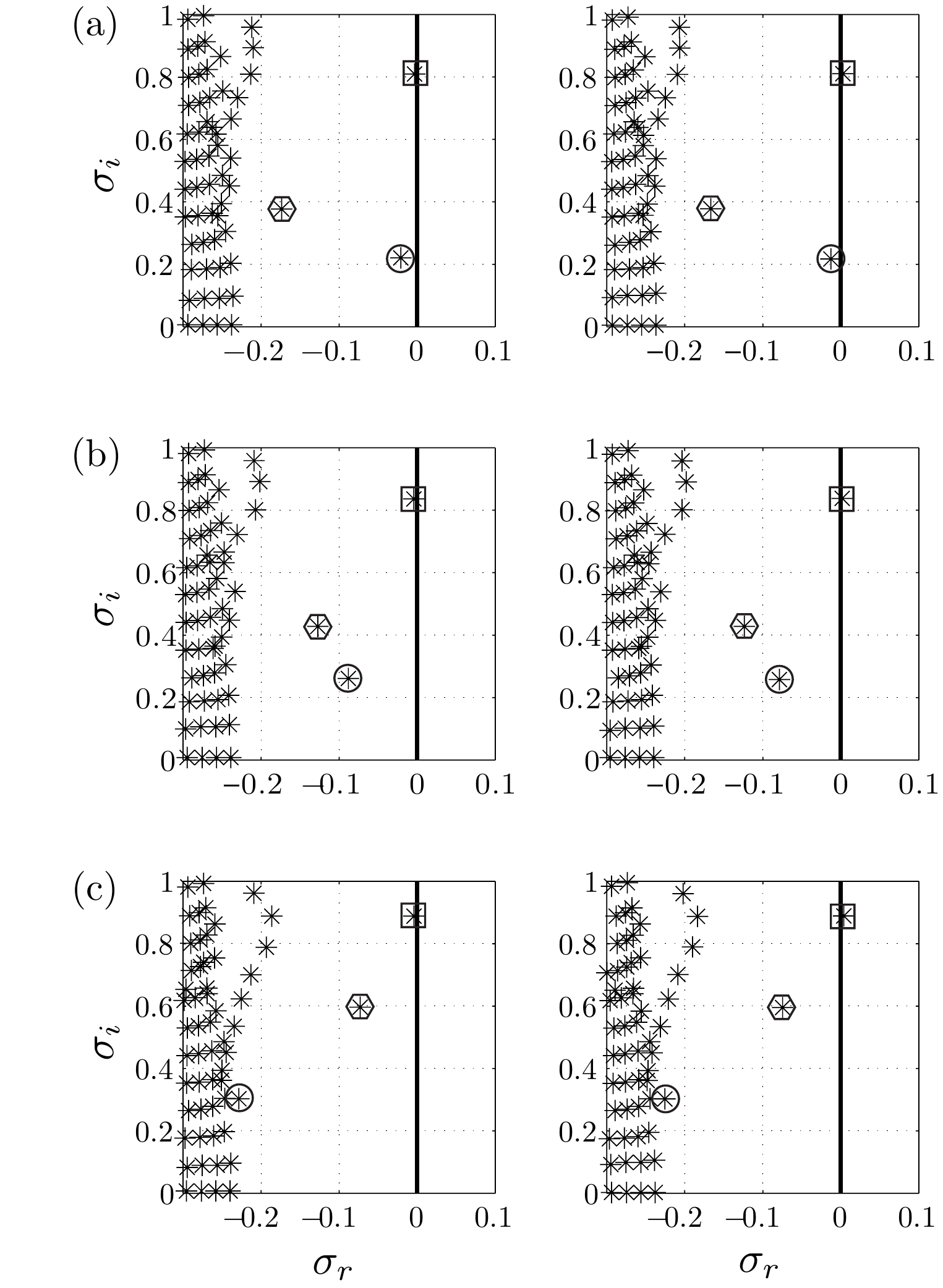}
\caption{Eigenvalue spectra for $\Rey=390$ (left column), $\Rey=400$ (right column) and $m=-1$ when (a) $\Omega=0.25$, (b) $\Omega=0.3$ and (c) $\Omega=0.4$.} \label{0p25_0p4}
\end{center}
\end{figure}
\begin{figure}
\begin{center}
\includegraphics[width=13cm]{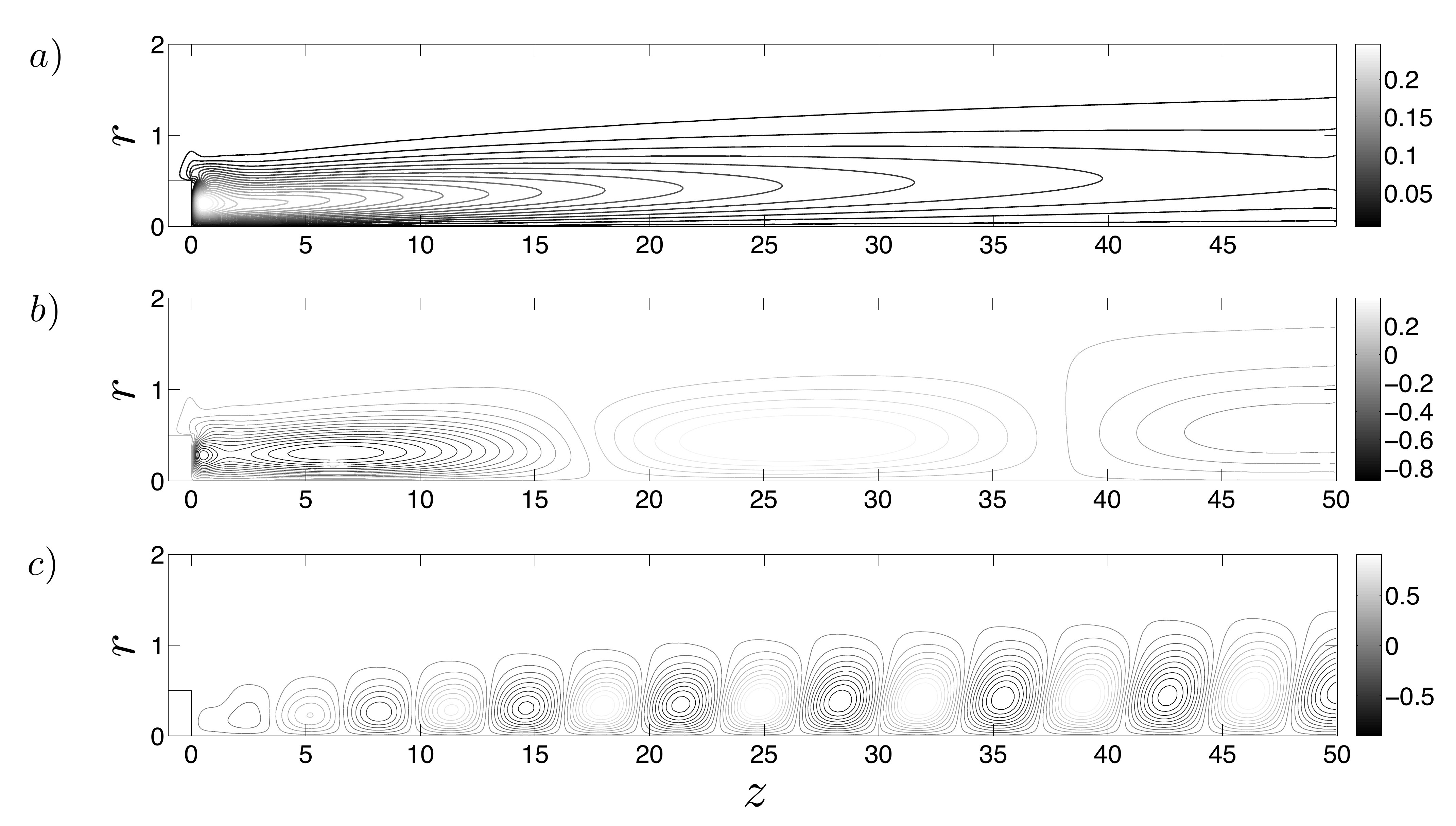}
\caption{Isocontours of the real part of the normalized axial velocity eigenfunction, $\R(\hat{w})/\|\hat{q}\|_{\infty}$, associated to the leading eigenvalue at (a) $\Omega=0$ and $\Rey=330$, (b) $\Omega=0.15$ and $\Rey=350$, and (c) $\Omega=0.25$ and $\Rey=400$.
\label{ef_leq_0p25}}
\end{center}
\end{figure}
\begin{table}
\centering
\begin{tabular}{c c c c c c c}
    $\Omega$  &  $\Rey_{c3}^{GLS}$  &  $\Str^{GLS}$  &  $\Rey_{c3}^{DNS}$  &  $\Str^{DNS}$ &
    $\epsilon_{\Rey_{c3}}(\%)$ &  $\epsilon_{\Str}(\%)$\\ \\
     $0.25$  &  395.78  &  0.1290  &  388.09  &  0.1288 & 1.9815 & 0.1553 \\
     $0.3$  &  397.10  &  0.1334  &  388.87  &  0.1321 & 2.1164 & 0.9841 \\
     $0.4$  &  395.34  &  0.1414  &  388.20  &  0.1417 & 1.8393 & 0.2117 \\
\end{tabular}
\caption{Critical Reynolds numbers for the destabilization of the HF mode, $\Rey_{c3}$, and Strouhal numbers obtained at $\Rey=400$, for several values of $\Omega$. Also shown are the values obtained by~\cite{Jimenez13} with direct numerical simulations (DNS). The relative errors, $\epsilon_{f}(\%)=|f_{GLS}-f_{DNS}|/f_{DNS}\times100$, being $f$ either $\Rey_{c3}$ or $St$, are also listed.}\label{Rec3}
\end{table}

According to previous numerical simulations, the nonlinear spiral frozen state (see figure~\ref{spiral_frozen}) prevails for $\Omega>\Omega_{c1}\simeq 0.232$, up to at least $\Omega=0.4$. In addition, the direct numerical simulations revealed that $\Rey_{c3}$ barely changes with $\Omega$. Both results are in agreement with the behaviour of the HF mode, as can be deduced from figure~\ref{0p25_0p4}, which shows the spectra for $\Omega=0.25,0.3$ and $0.4$ in the vicinity of the transition, and also from table~\ref{Rec3}. Figure~\ref{0p25_0p4} and table~\ref{Rec3} indicate that, for $0.25\leq \Omega \leq 0.4$, the HF mode is destabilized within a range of Reynolds numbers $390\leq \Rey\leq 400$. Table~\ref{Rec3} shows that, although the relative errors between GLS and DNS results are slightly larger than those obtained for $\Rey_{c1}$ (see table~\ref{Rec1}), they are nevertheless below $2.12\%$. Concerning the characteristic frequencies, table~\ref{Rec3}
displays the values of $St$ obtained at a slightly supercritical Reynolds number, $\Rey=400$, where it can be seen that the relative differences between the GLS and DNS results are below $1\%$. Moreover, figure~\ref{0p25_0p4} shows that the LF mode is stabilized as $\Omega$ increases at constant $\Rey$. At the same time, the MF mode becomes destabilized, although it is still stable at $\Omega=0.4$ (see figure~\ref{0p25_0p4}c).

Figure~\ref{ef_leq_0p25} shows the real part of the axial velocity eigenfunction, $\R(\hat{w})$, normalized with $\|\hat{q}\|_{\infty}$, associated to the leading eigenvalue obtained at three different slightly supercritical conditions, namely $(\Omega,\Rey)=(0,330),\,(0.15,350)$ and $(0.25,400)$. In the case without rotation, $\Omega=0$, figure~\ref{ef_leq_0p25}(a) shows the steady elongated structures, that lead to the SS state in the nonlinear regime~\citep{Fabre08,Bohorquez11}. However, when rotation is applied, the spatial structure of the wake changes qualitatively depending on the value of $\Omega$ compared with the critical value $\Omega_{c1}$. Indeed, for a value of the rotation parameter $\Omega=0.15<\Omega_{c1}$, figure~\ref{ef_leq_0p25}(b) reveals a pattern with a large but finite axial wavelength, corresponding to the LF eigenmode, that is responsible for the low-frequency frozen state (see figure~\ref{0p1}b). Let us point out that the frozen structures found in the direct numerical simulations
have a wavelength $\lambda_{DNS}\sim 39$ in the near wake, while the LF eigenmode provides $\lambda_{LF}\sim 43$. On the other hand, figure~\ref{ef_leq_0p25}c shows that, when $\Omega=0.25>\Omega_{c1}$, the HF eigenmode leads to a wake pattern with a much smaller characteristic axial wavelength of $\lambda_{HF}\sim 6.4$, while the numerical results provide a similar value of $\lambda_{DNS}\sim 6.6$ for the corresponding frozen spiral state.

\begin{figure}
\begin{center}
\includegraphics[width=12cm]{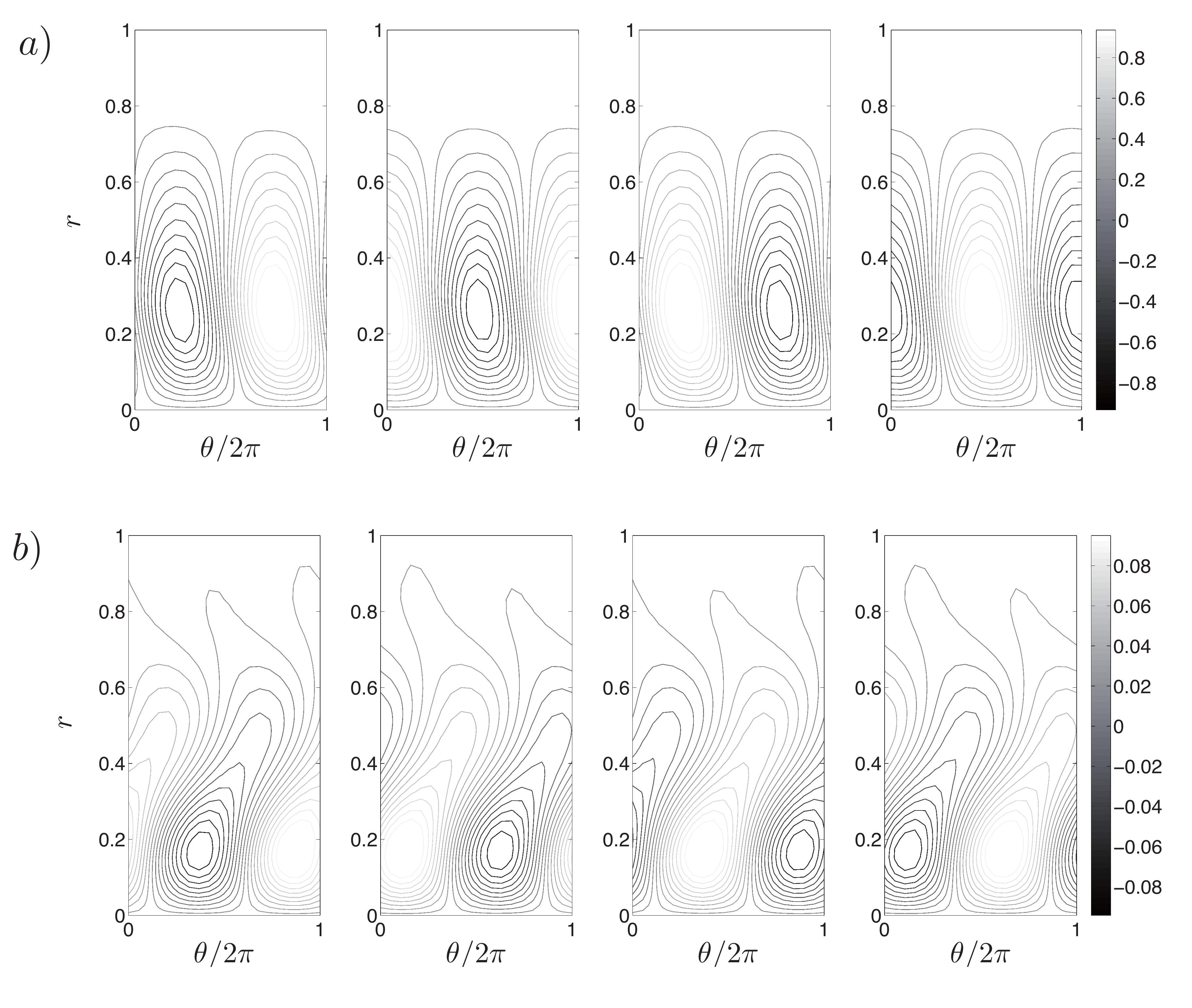}
\caption{Snapshots of the temporal evolution of the contours of the normalized axial velocity disturbance, $\Re(w')$, defined by equation~(\ref{phase}), associated to the leading eigenmode, evaluated at $z=1$, for (a) $\Omega=0.15$ and $\Rey=350$, (b) $\Omega=0.25$ and $\Rey=400$. The time interval between snapshots is $T/4$, being $T=2\pi/\sigma_i$.}
\label{temp_evo_w}
\end{center}
\end{figure}

A fundamental feature of the flow in the supercritical regime is the co-rotating or counter-rotating nature of the structures emerging from the bifurcation. Figure~\ref{temp_evo_w} depicts one period of the temporal evolution of the normalized axial
velocity disturbance,

\begin{equation}
\R(w')=\frac{|\hat{w}|\,e^{\sigma_r t}}
{\|\hat{\mathbf{q}}\|_{\infty}}\cos\left[\sigma_i t+m\theta+
\arctan\left(\frac{\Im(\hat{w})}{\Re(\hat{w})}\right)\right],\label{phase}
\end{equation}
for two different cases dominated either by the LF mode (figure~\ref{temp_evo_w}a, where $\Omega=0.15$ and $\Rey=350$) or the HF mode (figure~\ref{temp_evo_w}b, with $\Omega=0.25$ and $\Rey=400$). It is deduced from figure~\ref{temp_evo_w} that, in both cases, the disturbances rotate in the positive azimuthal direction, i.e. with the angular velocity vector oriented in the positive $z$-direction. It is worth pointing out that the same conclusion can be inferred directly from the following argument: the infinitesimal time increment of the disturbance phase, $\sigma_i dt+m d\theta$, is zero in a reference frame that rotates with the angular velocity of the structures, namely ${\rm d}\theta/{\rm d}t=-\sigma_i/m$. Therefore, a given global eigenmode corresponds to a counter-rotating structure if $\sigma_i$ and $m$ have the same sign, while it is co-rotating if $\sigma_i$ and $m$ have opposite signs, as in the case
of the LF, MF and HF modes found in the present work.
\begin{figure}
\begin{center}\hspace{-1cm}
\includegraphics[width=12cm]{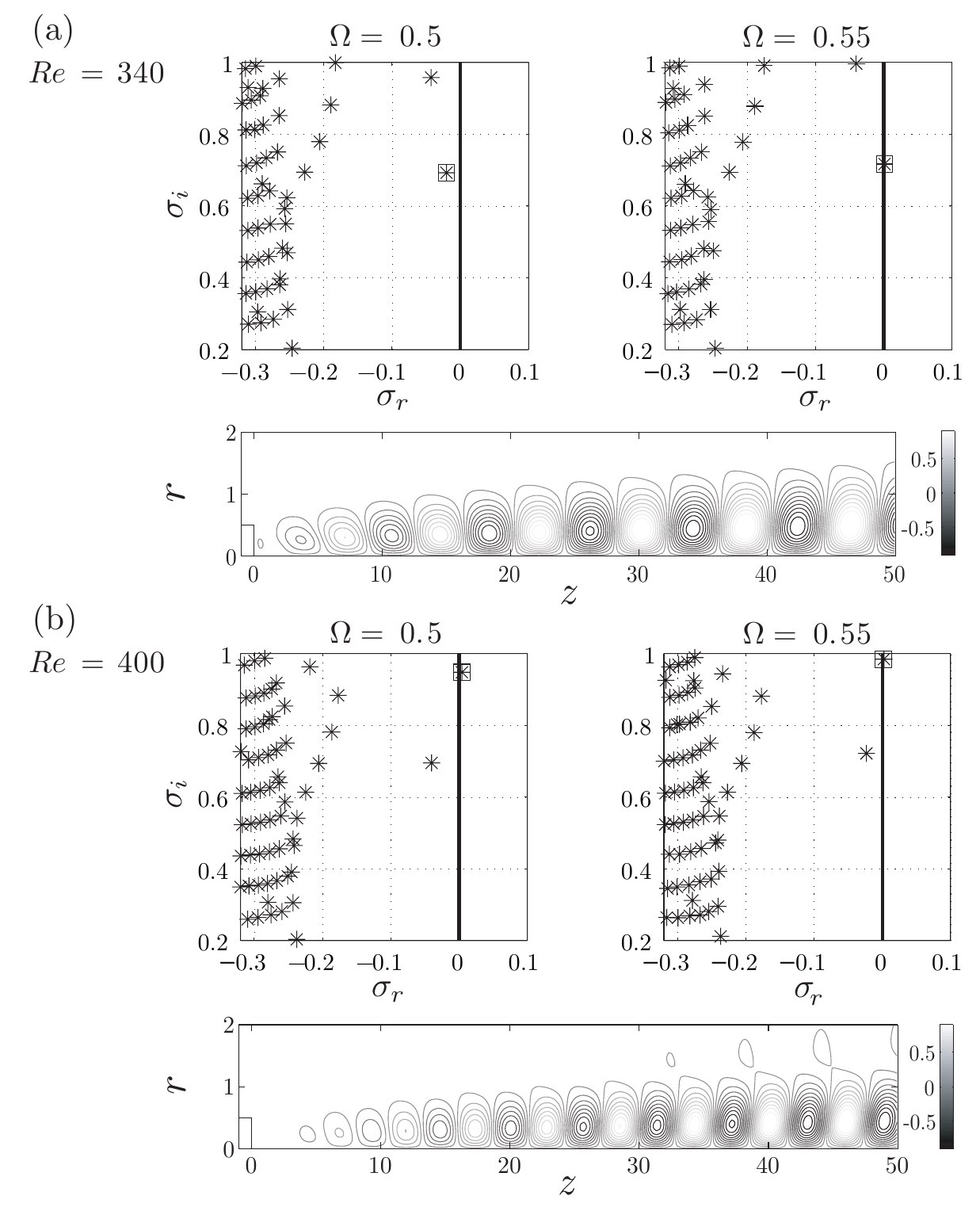}
\caption{Eigenvalue spectra for $\Omega=0.5$ and $\Omega=0.55$, with (a) $\Rey=340$ and (b) $\Rey=400$. The contour plots display isolines of $\R(\hat{w})/\|\hat{\mathbf{q}}\|_{\infty}$ for (a) the MF eigenmode at $\Omega=0.55$ and $\Rey=340$, and (b) the HF eigenmode at $\Omega=0.55$ and $\Rey=400$.\label{0p5_0p55}}
\end{center}
\end{figure}

\subsection{Stability characteristics for $0.4<\Omega\leq1$}\label{SectRESULTSb}

Given that the stability analysis has been successfully applied to understand the transition regimes described in \S~\ref{SectRESULTS0}  in the range of rotation parameters $\Omega\leq 0.4$, the present section is devoted to report the results of an extended investigation, where the rotation parameter is increased up to $\Omega=1$.

\begin{figure}
\begin{center}
\includegraphics[width=7cm]{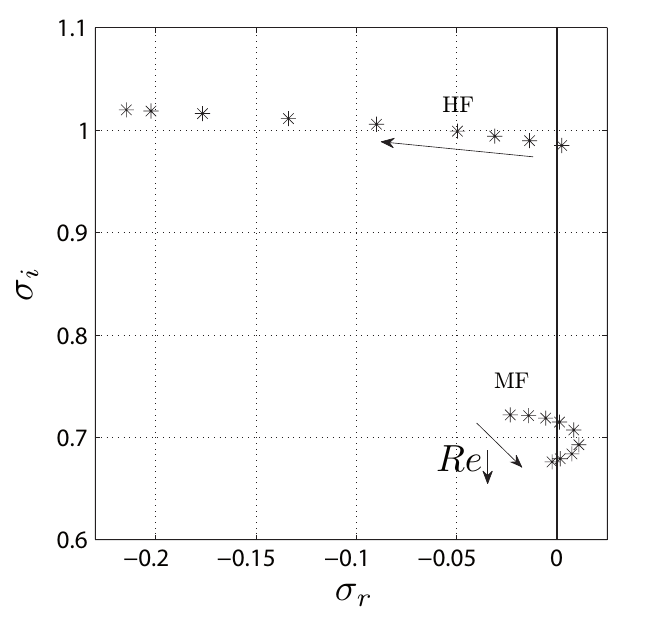}
\caption{Eigenvalue spectra obtained for a body rotating with a velocity $\Omega=0.55$ at $\Rey=400,380,360,340,300,270,240,220$ and $210$. Arrows indicate the direction of decreasing $\Rey$.\label{0p55}}
\end{center}
\end{figure}

Figure~\ref{0p5_0p55} shows that, as $\Omega$ increases, the axisymmetric state may be destabilized by either the MF or the HF eigenmodes, depending on the value of $\Rey$. In particular, although at $\Rey=400$ the HF mode is still responsible for the transition found by increasing $\Omega$, the MF eigenvalue is first destabilized at $\Omega\simeq 0.55$ at a smaller value of $\Rey=340$. The latter result is representative of the transition scenario at high enough values of $\Omega$ and low enough values of $\Rey$. However, the behaviour of the MF mode at constant $\Omega$ and varying $\Rey$ is more complicated than those found for the LF and HF modes. Indeed, from figure~\ref{0p55} it is deduced that, for $\Omega=0.55$, the MF mode is unstable within a range of Reynolds numbers $220 \leq\Rey\leq 340$.
The real part of the normalized axial velocity associated to the MF eigenmode, $\R(\hat{w})/\|\hat{\mathbf{q}}\|_{\infty}$, obtained for $\Omega=0.55$ and $\Rey=340$ is displayed in figure~\ref{0p5_0p55}(a), where it can be seen that the spatial structure is similar to that of the HF mode, shown in figure~\ref{0p5_0p55}(b) for $\Omega=0.55$ and $\Rey=400$.

\begin{figure}
\begin{center}
\includegraphics[width=12cm]{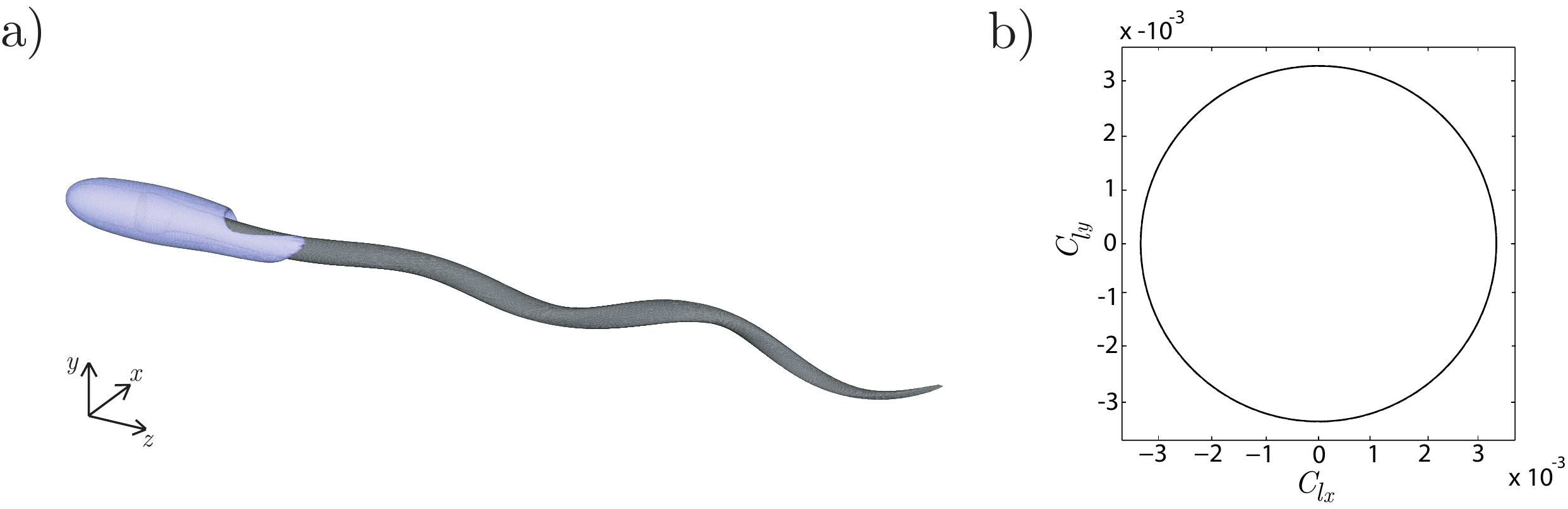}
\caption{(a) Isosurfaces of streamwise vorticity, $\varpi_{z}=0.4$ (dark-coloured contour), and $\varpi_{z}=-0.4$ (light-coloured contour), and (b) lift phase diagram, for $\Omega=0.55$ and $\Rey=340$.\label{NS0p55}}
\end{center}
\end{figure}

The behaviour of the MF mode in the nonlinear regime was also studied by performing additional unsteady three-dimensional simulations for $\Omega=0.55$ and $\Omega=0.6$, and several values of $\Rey$. As an example, figure~\ref{NS0p55} shows the results obtained when $\Omega=0.55$ and $\Rey=340$. In particular, the isocontours of axial vorticity displayed in figure~\ref{NS0p55}(a) reveal the presence of spiral structures, while the phase diagram built with the two components of the lift coefficient, $[C_{l_x}(t),C_{l_y}(t)]$, depicted in figure~\ref{NS0p55}(b), demonstrates that the wake is in a frozen regime. Here, we have defined the vector lift coefficient $\mathbf{C}_l=8\,\mathbf{F}_l/[\rho w_{\infty}^2\, D^2( \pi +4)]$, where $\mathbf{F}_l$ is the dimensional lift force, being $C_{l_x}$ and $C_{l_y}$ the projections of $\mathbf{C}_l$ on the $x$- and $y$-axes, respectively and $D^2( \pi/4 +1)$ the lateral area of the body. Moreover, the numerical simulations show that the wake structures rotate in the same
direction as the body, but with a different angular velocity. In view of these features, it seems reasonable to denote the newly found nonlinear regime as \emph{Medium-Frequency Spiral Frozen mode}. A noteworthy feature of this new wake state is the fact that it leads to a much higher lift force than in the high-frequency spiral frozen state, as can be deduced from the results shown in table~\ref{table_0p55_0p6}. Note the good quantitative agreement found between the frequencies predicted by the numerical simulations and the global stability analysis, with relative errors $\epsilon_{St}<4\%$. Concerning the periodicity of the axial structures, the results from direct numerical simulations and global linear stability analysis respectively provide local wavelengths $\lambda_{DNS}\sim 8$  and $\lambda_{GLS}\sim 7.6$ in the near wake.

\begin{table}
\centering
\begin{tabular}{c c c c c c c c c c c}
     $\Omega$ & &   & $\Rey=340$ &  &  & &  & $\Rey=400$& $$ & $$  \\ \cline{3-6} \cline{8-11}
     & &  $\Str_{DNS}$  &  $\Str_{GLS}$  &  $\epsilon_{St}(\%)$  &  $C_l$   &  &
    $\Str_{DNS}$  &  $\Str_{GLS}$  &  $\epsilon_{St}(\%)$ &  $C_l$ \\ \\
     0.55 &  &  0.1106  &  0.1138  &  2.768  &  0.0032 & & 0.1583 & 0.1567 & 0.960 & 0.0008 \\
     0.6  &  &  0.1134  &  0.1175  &  3.527  &  0.0049 & & 0.1621 & 0.1626 & 0.297 & 0.0010 \\
\end{tabular}
\caption{Strouhal numbers obtained through global stability analysis, $\Str_{GLS}$,  and three-dimensional numerical simulations, $\Str_{DNS}$, and lift coefficient, $C_l$, determined numerically, for $\Omega=0.55$ and $\Omega=0.6$, at $\Rey=340$ and $400$. The relative error, $\epsilon_{St}(\%)=|St_{GLS}-St_{DNS}|/St_{DNS}\times100$, is also shown.
\label{table_0p55_0p6}}
\end{table}

\begin{figure}
\begin{center}
\includegraphics[width=13cm]{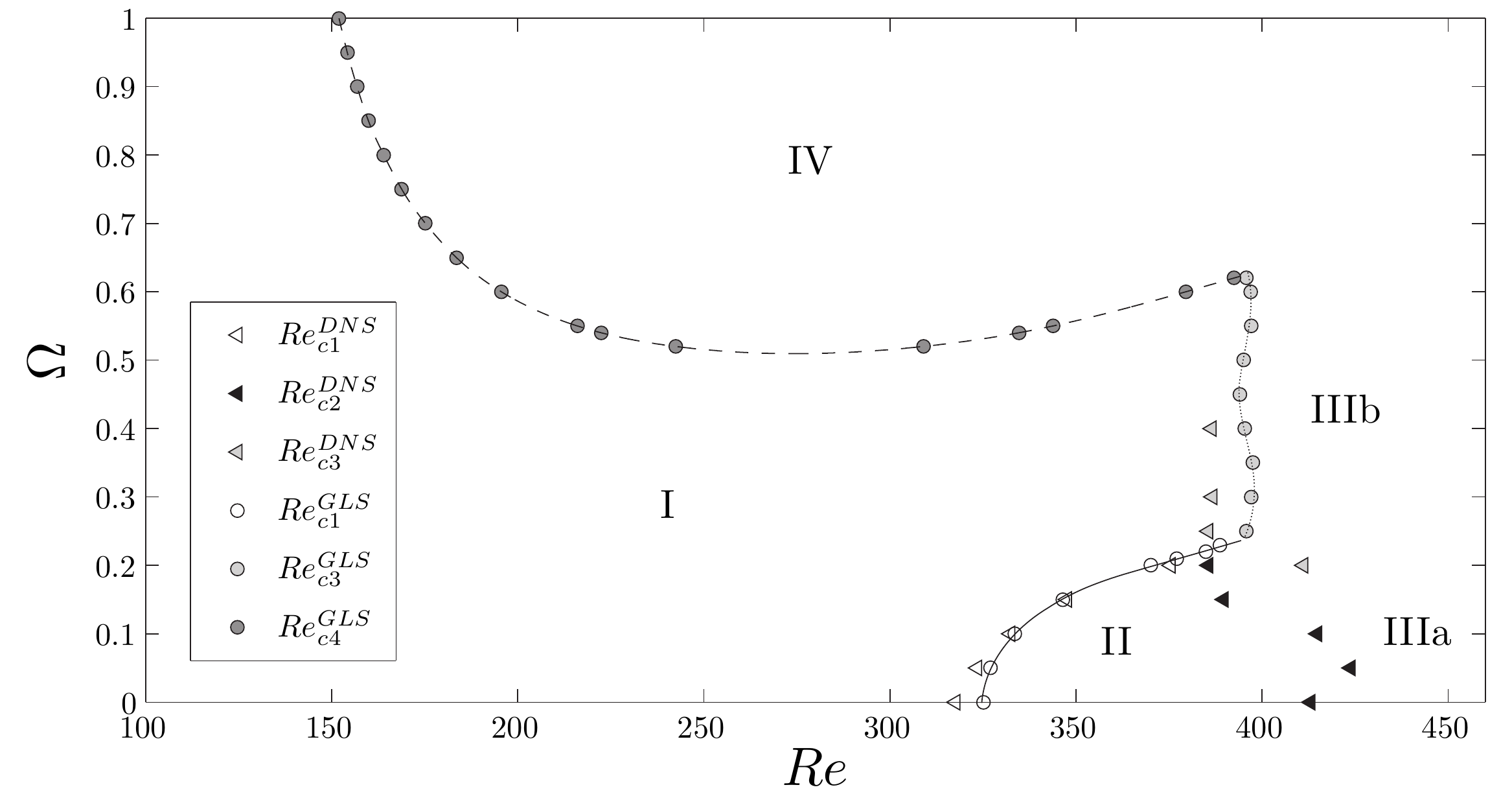}
\caption{Bifurcation diagram in the ($\Omega$,$\Rey$) parameter plane, according to the results of numerical simulations (DNS) and global stability analysis (GLS). The different wake states correspond to the following regions: I axisymmetric, II frozen, IIIa spiral unsteady, IIIb high-frequency spiral frozen and IV medium-frequency spiral frozen.\label{transitions}}
\end{center}
\end{figure}

The neutral curve associated to the MF mode was tracked within the ranges of Reynolds number and rotation parameter $0\leq \Rey\leq 450$ and $0 \leq\Omega\leq 1$, respectively, to complete the bifurcation diagram in the ($\Omega,\Rey$)-parametric plane, shown in figure~\ref{transitions}. This figure also includes the critical conditions reported at low values of $\Omega$ by~\cite{Jimenez13} (triangles in figure~\ref{transitions}). The diagram displays the four unstable nonlinear regimes observed around the axisymmetric state (Region I), namely: the \emph{Frozen} State (see figure~\ref{0p1}b), within Region II, associated to the destabilization of the LF linear mode; the \emph{High-Frequency Spiral Frozen} State (see figure~\ref{spiral_frozen}), within Region IIIb, associated to the destabilization of the HF linear mode; the \emph{Medium-Frequency Spiral Frozen} State (see figure~\ref{NS0p55}a), within Region IV, related to the destabilization of the MF linear mode; and the \emph{Spiral Unsteady} State (see figure~\ref{0p1}c) in Region IIIa. The latter state seems to be the outcome of the nonlinear interaction between the linear LF and HF eigenmodes, which are unstable within Region IIIa. Indeed, their frequencies are similar to those revealed by the force and velocity spectra obtained from direct numerical simulations of the nonlinear Spiral Unsteady State close to criticality.

The global linear stability analysis performed has provided the limits of the axisymmetry (Region I) at high values of $\Omega$, showing that $\Rey_{c3}$ is approximately independent of $\Omega$ (in the range $0.232\lesssim\Omega\lesssim 0.625$). Moreover, the unstable Region IV exists only for $\Omega\geq\Omega_{min}\simeq 0.509$ approximately, for which the critical Reynolds number is $\Rey_{c4}(\Omega_{min})\simeq 274$.
Note from figure~\ref{transitions} that $\Rey_{c4}$ increases with $\Omega$ for $\Rey >\Rey_{c4}(\Omega_{min})$, while the value of $\Rey_{c3}$ remains almost constant, both marginal curves intersecting at $\Omega_{c2}\simeq 0.625$ and $\Rey_{c4}(\Omega_{c2})\simeq 396$. On the other hand, $\Rey_{c4}$ decreases as $\Omega$ increases for $\Rey < \Rey_{c4}(\Omega_{min})$, apparently reaching an asymptote for $\Omega>1$, so that Region IV could be bounded by a region of axisymmetric flow (Regime I) at low enough values of $\Rey$ independently of the value of $\Omega$. Besides, in the region where both the HF and MF modes coexist, we expect the existence of a new unsteady regime due to the nonlinear interaction between both global modes, although further numerical simulations will be required to clarify this issue.

\begin{figure}
\begin{center}
\includegraphics[width=10cm]{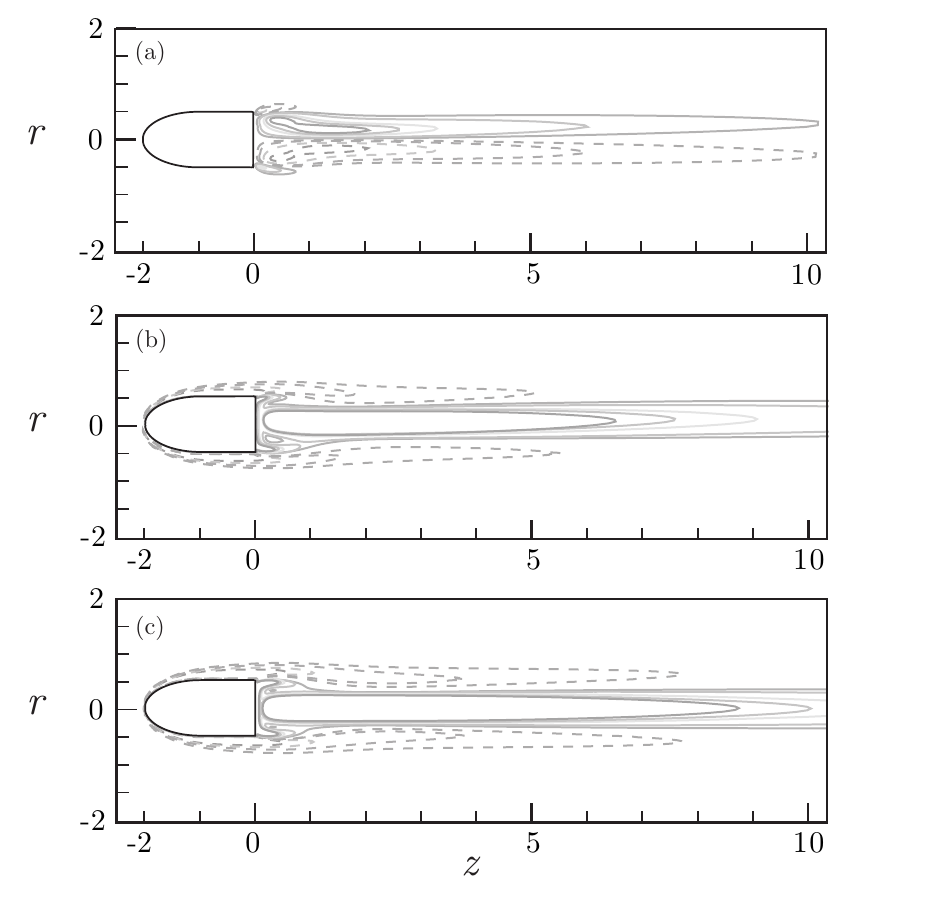}
\caption{Contours of axial vorticity, $\varpi_z$, obtained from DNS results at $\Rey=340$ for: (a) $\Omega=0$, (b) $\Omega=0.1$ and (c) $\Omega=0.15$. Vorticity levels: $\pm0.3$, $\pm0.25$, $\pm0.2$, $\pm0.1$, $\pm0.05$ (dashed lines represent negative values).\label{Ax_vort}}
\end{center}
\end{figure}

\subsection{Physical interpretation of results}\label{SectRESULTSc}
The unstable modes described above are the outcome of the modification, through the centrifugal force and the azimuthal shear, of the axial shear modes associated with the wake of non-rotating bodies, and their behaviour and onset could be explained through the competition among these effects. According to~\cite{Johnson99}, the axisymmetry-breaking bifurcation that takes place in the wake of a sphere is the consequence of an azimuthal instability, whereby an azimuthal pressure gradient promotes the flow between the upper and the lower vortex cores in the near wake~\citep[see figure12 in][]{Johnson99} and opens the recirculation bubble. Consequently, the fluid inside the recirculating region is released, giving rise to axial vorticity tails in the wake. Due to the planar symmetry of the SS mode, the flow between cores occurs in both, the positive and the negative azimuthal directions. When rotation is applied in our bullet-like body, an axisymmetric linear distribution of azimuthal velocity is imposed at the body base. Therefore, the flow induced between the two vortex cores is reduced in the azimuthal direction opposite to the spin, decreasing the intensity of the axial vorticity thread in this direction, and promoting the one rotating with the base. This effect can be observed in figure~\ref{Ax_vort}, where axial vorticity $\varpi_z$ contours from numerical simulations are plotted for $\Rey=340$ and increasing values of $\Omega$. In fact, figure~\ref{Ax_vort} reveals that the planar symmetry associated with the SS mode at $\Omega=0$ (figure \ref{Ax_vort}a) is broken when rotation is applied, due to the injection of positive vorticity from the body base (see figure \ref{Ax_vort}b for $\Omega=0.1$) that promotes the positive tail and weakens the negative one. Additionally, a shroud of negative $\varpi_z$ develops from the nose when the body spins. This effect becomes stronger as the angular velocity of the body increases, until eventually the axisymmetry is retrieved at a critical value of $\Omega$, what is indeed generally observed for $\Rey<400$. For instance, at $\Rey=340$ the stabilization occurs at $0.1<\Omega<0.15$, as is evidenced by figure~\ref{Ax_vort}c.

Following the results of~\citet{Ghidersa00} for the sphere, which demonstrated that the bifid wake (SS mode) is a combination of the axisymmetric base flow plus the linear $m=1$ mode (with a weak nonlinear correction), an explanation for the retrieving of axisymmetry could also be elaborated in terms of global modes properties. Thus, the weakening of the (now rotating) bifid topology with increasing spin velocity might be seen as the combined effect of a stronger azimuthal base flow velocity and a weaker amplitude of the $m=-1$ perturbation velocity~\citep[in line with figure 8 in][]{Ghidersa00}. On the other hand, the weakening of the amplitude of the dominant unstable mode (LF mode) as $\Omega$ grows, should lead to a better prediction of Strouhal numbers for the subdominant mode (HF mode),  as suggested by the results of~\cite{Tomboulides00}. Indeed, these authors discussed, in terms of the dependence of $\Str$ with $\Rey$, the errors associated with the use of an axisymmetric base flow to predict the destabilization of the subdominant linear mode beyond the first axisymmetry-breaking bifurcation (see figure 12 in their paper). This allowed~\citet{Tomboulides00} to interpret the results of the global stability of the flow around a sphere performed by~\cite{Natarajan93}, highlighting the fact that the subdominant linear mode must become unstable for the onset of vortex shedding to take place. In the case at hand, due to the proximity of both bifurcations at $\Rey_{c1}$ and $\Rey_{c2}$ in the vicinity of $\Omega\simeq 0.232$, not only the prediction of $\Str$ for the subdominant HF mode is satisfactory, but also for the second bifurcation~\citep[note that for $\Omega=0$ there is an important underprediction of $\Str$ and an overprediction of $\Rey_{co}$ for the RSP mode by the stability analysis, according to][]{Bohorquez11}. Thus, under these conditions the performance of the stability analysis beyond the axisymmetry breaking bifurcation improves with respect to the case without rotation, similarly to what occurs for the wake of a disk~\citep{Natarajan93}.
\begin{figure}
\begin{center}
\includegraphics[width=11cm]{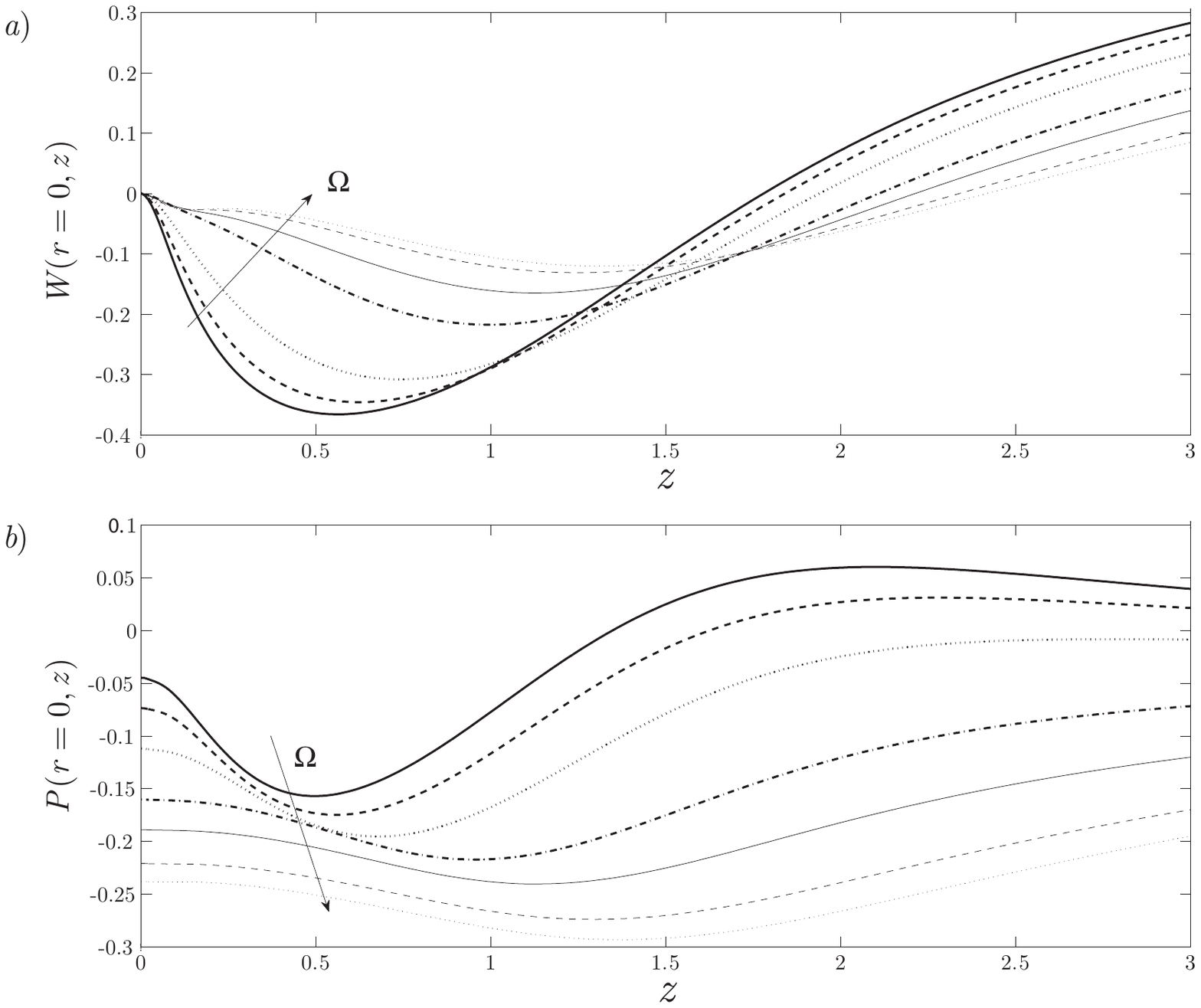}
\caption{Effect of the rotation parameter $\Omega$ on the axisymmetric base flow (a) centerline axial velocity, $W(r=0,z)$ and (b) centerline pressure, $P(r=0,z)$, for $\Omega=0,0.15,0.25,0.4,0.5,0.6$ and $0.65$, when $\Rey=400$.\label{BF_analysis}}
\end{center}
\end{figure}
To shed some light on the physical mechanisms governing the stability of the axisymmetric state, figure~\ref{BF_analysis} shows the downstream evolution of the axial velocity and the static pressure at the axis, $W(r=0,z)$ and $P(r=0,z)$ respectively, for several values of $\Omega$ at $\Rey=400$. Note that the latter value of the Reynolds number is close to the marginal conditions for a wide range of rotation parameters. From the results of figure~\ref{BF_analysis}(a) it is deduced that, as $\Omega$ increases, the recirculation region enlarges and the maximum backflow velocity decreases, leading to a decrease in the radial gradient of axial velocity, similar to the effect of base bleed~\citep{Sevilla04}. This resulting reduction of axial shear might contribute to stabilize the wake, since, as shown by~\citet{Monkewitz88}, reverse flow is known to promote local absolute instability for non-rotating cases~\cite[which is in fact connected to the existence of global instability, as demonstrated for the sphere by][]{Pier08}, and it could have a similar effect when rotation is applied. Moreover, this reduction of axial velocity gradients at high $\Omega$ renders the flow more parallel. Consequently, in line with the results obtained at high bleed coefficients for the wake of the body without rotation~\citep{Bohorquez11}, where the slenderness of the flow justifies the use of local parallel base flow, a local stability analysis at several downstream locations from the base using a quasi-parallel flow approximation, might characterize approximately the transition between Region I and IV defined in figure~\ref{transitions}. On the contrary, at low $\Omega$, the deviation from parallel flow in the near wake would surely lead to discrepancies between the critical parameters determining the occurrence of local absolute instability and the onset of global self-sustained oscillations, as shown for the cylinder~\citep{Monkewitz88b} and the non-rotating sphere~\citep{Pier08}. At the same time, figure~\ref{BF_analysis}(a) shows an incipient inflection point in the centerline axial velocity near the body (at $z=0.25$ approximately) that, at higher angular velocities, could lead to the formation of a second recirculation bubble, resembling the \emph{vortex breakdown} phenomenon, although this hypothesis should be explore in the future. The results depicted in figure~\ref{BF_analysis}(b) also show that the  base pressure decreases as $\Omega$ increases, leading to an increase in the drag force,  while the axial location of the minimum pressure is displaced downstream, reflecting the enlargement of the recirculation region.

Let us finally discuss our results in the light of related studies that deal with the stability of rotating flows, as well as with the theory of equivariant bifurcations. The introduction of body rotation results in an axisymmetry breaking flow, through supercritical Hopf bifurcations, associated with the LF, MF and HF eigenmodes found in the present work. Note that, this type of bifurcating behaviour has been previously reported in other configurations like enclosed rotating flows~\citep{Lopez01}, or a counter-rotating Taylor-Couette flow~\citep{Langford88,Crawford91}. This bifurcating scenario is a consequence of the invariance of the base flow under arbitrary rotations around the axis, constituting a $SO(2)$-symmetric configuration which, as revealed by the theory of equivariant bifurcations, leads to \emph{rotating waves}, i.e. solutions that are steady in a reference frame that rotates with an appropriate angular velocity~\citep{Ruelle73}. The results reported herein, clearly demonstrate that the low-frequency, the medium-frequency and the high-frequency frozen wakes, that appear as a consequence of the destabilization of the LF, MF and HF eigenmodes respectively, indeed match the characteristics of \emph{rotating wave} states. On the other hand, once the next-to-leading eigenmode becomes unstable, the wake adopts a quasi-periodic motion with two characteristic frequencies through another Hopf bifurcation, featuring an intrinsic unsteadiness that leads to vortex shedding in the wake, similar to the \emph{modulated wave} states that appear in Taylor-Couette Flow~\citep{Golubitsky00} and other enclosed swirling flows~\citep{Blackburn02}.

\section{Conclusions}\label{SectCONCLUSIONS}

We have studied the global linear stability of the axisymmetric flow around a spinning bullet-shaped body of length-to-diameter ratio $\ell=2$, as a function of the Reynolds number, $\Rey$, and of the rotation parameter $\Omega$, which have been varied in the ranges $\Rey<450$ and $0\leq\Omega\leq 1$. The stability analysis has provided a detailed characterization of the unstable modes that appear in the wake and of its corresponding bifurcated states.

First, it can be pointed out that axial shear drives generally the flow instability for low and moderate values of $\Omega$, as suggested by the fact that only helical ($m=\pm1$) global modes are destabilized for non-rotating bodies \citep[][among others]{Monkewitz88,Natarajan93} and rotating body wakes. This common instability mechanism is more pronounced at very low values of $\Omega$ (e.g. $\Omega<0.232$ for the geometry considered herein), where, despite the unsteadiness, the flow topology resembles that of the SS mode for non-rotating bodies. However, rotation provides with a more complex and richer stability picture when $\Omega$ increases. Thus, in summary, three different co-rotating eigenmodes with helical azimuthal symmetry have been identified, each of them dominating the instability of the axisymmetric state in different regions of the $(\Rey,\Omega)$ parameter plane. The first eigenmode is a \emph{Low-Frequency mode}, destabilized for $\Omega\leq 0.232$ and $\Rey>\Rey_{c1}(\Omega)$, whose associated eigenfunction features structures with large axial wavelength, and is responsible for the nonlinear \emph{Frozen State} described in~\cite{Jimenez13}. In the range of rotation parameters $0.232<\Omega<0.62$, a different \emph{High-Frequency mode} becomes destabilized for $\Rey>\Rey_{c3}(\Omega)$, with a much smaller axial wavelength, that gives rise, in the nonlinear regime, to an \emph{Spiral Frozen wake}. Finally, a new \emph{Medium-Frequency mode} has been found, that is destabilized for $\Omega\geq 0.51$ within a range of Reynolds numbers that widens with increasing $\Omega$. The nonlinear counterpart of the medium-frequency mode has also been studied by means of unsteady three-dimensional numerical simulations, revealing a development into a co-rotating \emph{frozen spiral} topology of the wake, similar to that associated with the high-frequency mode. Therefore, this new wake state can be described as a \emph{Medium-Frequency Spiral Frozen mode}. All these nonlinear regimes existing around the axisymmetric region in the parametric ($\Omega,\Rey$)-map have been found to behave as \emph{rotating waves} that, according to the theory of equivariant bifurcations, is a consequence of the $SO(2)$-symmetry of the base flow.

The stability picture described in this work for the bullet-shaped body differs considerably from that of the sphere~\citep{Kim02,Pier13}. Indeed, in contrast to what happens in the flow around a sphere, where streamwise rotation destabilizes the axisymmetric state, in the present case rotation acts as a control mechanism for moderate values of $\Omega$. Moreover, no evidences of a mode similar to the medium-frequency one reported herein has been observed in the wake of a rotating sphere~\citep{Pier13}. These differences are related to the effect of the body base on the flow field that, at low values of $\Omega$, enhances the generation of vorticity in a preferred sense, decreasing the intensity of the axial vorticity thread of opposite sign and contributing to retrieve the axisymmetry. However, the presence of the new unstable medium-frequency mode at higher values of $\Omega$ limits this stabilizing effect of rotation. An analysis of the base flow has shown that, when rotation is applied, the value of the recirculating axial velocity decreases at the centerline, which fosters the stabilization of the wake at low values of $\Omega$; but at high values of rotation parameter, the centrifugal mechanism increases the magnitude of the radial velocity at the base of the body, what might destabilize the incident stream at the rear edge, and consequently the wake. Additionally, at high $\Omega$, an inflection point begins to develop at the centerline axial velocity distribution, that eventually could lead to the break-up of the recirculation bubble into two separate pieces, creating a flow topology that may resemble that of a vortex breakdown. This possibility, along with a deeper study on the influence of the body base in rotating blunt-based bodies (e.g. disks or cylinders) on the stability properties of the wake, arise as interesting topics for future work, that could shed more light on the complicated dynamics of swirling wakes past spinning bluff bodies.

On the other hand, some comments must be made on the controllability of the flow. In an initial attempt to evaluate the sensitivity of the flow to local modifications of the linear Navier-Stokes operator (useful for passive control strategies implementation), it has been observed that, at low values of $\Omega$, slight base flow changes within the recirculation bubble, for instance by means of modifications in the boundary condition at the base (e.g. adding some bleed), lead to important shift in the growth rate of the spectra dominant eigenvalues. Consequently, we presume that the ``wakemaker'' is located within the recirculation bubble, in line with what has been observed for other axisymmetric bodies without rotation, such disks or spheres \citep{Meliga09}. This sensitivity to modifications in the base boundary condition seems to be smaller for high $\Omega$, so that it is likely that the wakemaker location changes with the spin value. However, further investigations are needed in the future to perform the identification of regions of large sensitivity.

Finally, the transition diagram obtained by means of the global linear stability analysis (figure~\ref{transitions}) shows good quantitative agreement with the three-dimensional numerical simulations. In fact, relative errors found in the critical Reynolds numbers are lower than $2.12\%$ for $\Omega\leq 0.4$, while the relative differences in the characteristic frequencies lie below $5.1\%$ for all $\Omega$ and marginally unstable $\Rey$ investigated. This result is interesting, not only from the practical point of view, in terms of computational efficiency, but also from the physical point of view, since it can be connected to the fact that the present flow is strongly non-parallel in the near wake and the convective non-normality might be moderate, which would minimize the risk of transient growth and noise-amplifier dynamics of the stable flow~\citep{Chomaz05}. Consequently, global linear stability analysis stands out as a powerful tool that predicts and properly characterizes the physics of the wake behind the spinning body around the stable region.

\begin{acknowledgments}
This work was supported by the Spanish MINECO under projects \#DPI2011-06624-C02 and
\#DPI2011-06624-C03, by Junta de Andaluc\'ia thanks to project P10-TEP5702, and
by University of Ja\'en under project UJA-2010-12-60.
\end{acknowledgments}

\begin{appendix}

\section{Numerical methods}\label{sec:numerics}

\subsection{Base flow computation}
The base flow was obtained solving equations~\eqref{BF1}-\eqref{BF2} with the finite-volume code ANSYS Fluent\textsuperscript{\textregistered} within the axisymmetric domain depicted in figure~\ref{scheme}, consisting of an inlet hemisphere with radius  $10D$, followed by a cylinder of length $50D$. Note that the computational domain is similar to that used in~\cite{Jimenez13}, where the domain was proven to be adequate to perform unsteady three-dimensional numerical simulations. The boundary conditions set in the axisymmetric simulations were
\begin{gather}
\mathbf{U}(\mathbf{x})=(0,0,w_{\infty}),~~~\mathbf{x}~\in~\Sigma_i, \\
\mathbf{U}(\mathbf{x})=(0,\omega r,0),~~~\mathbf{x}~\in~\Sigma_w, \\
\mathbf{n}\cdot\mathbf{U}(\mathbf{x})=0,~~~\mathbf{x}~\in~\Sigma_f, \\
P(\mathbf{x}) \, \mathbf{n}- \frac{1}{\Rey} \, \mathbf{n} \cdot \nabla
\mathbf{U}(\mathbf{x})=0,~~~\mathbf{x}~\in~\Sigma_o,\\
U(\mathbf{x})=V(\mathbf{x})=\frac{\partial
W}{\partial r}=\frac{\partial P}{\partial
r}=0,~~~\mathbf{x}~\in~\Sigma_a.
\end{gather}
\begin{table}
\centering
\begin{tabular}{c c c c c c c}
Mesh & $N$ & $V(r=0.5,~z=1)$ & $\text{GCI}_{j+1,j}(\%)$ & $W(r=0.5,~z=5)$ & $\text{GCI}_{j+1,j}(\%)$\\ \\
 $\#1$ & $103700$ & $0.03774$ & $0.23467$ &  0.80607 & 0.03908\\
 $\#2$ & $52000$ & $0.03771$ & $0.48076$ & 0.80618 & $0.07668$\\
 $\#3$ & $26200$ & $0.03764$ & $-$ & $0.80640$ & $-$ \\
\hline
$n$ & $-$ & $2.067$ & $-$ & $2.000$ & $-$ \\
\end{tabular}
\caption{The Grid Convergence Index for the basic flow, $\text{GCI}_{j+1,j}$, for values of $\Omega=0.1$ and $\Rey=330$, obtained using $V(r=0.5,z=1)$ and $W(r=0.5,z=5)$. Here the grid refinement ratio is $\alpha=\sqrt{2}$.}\label{GCI_BF}
\end{table}

The axisymmetric conditions at $\Sigma_a$ were obtained after avoiding the indeterminate axial terms through $L'H\hat{o}pital's$ $rule$ \citep{Gerritsma00}. Equations~\eqref{BF1}-\eqref{BF2} were discretized with second-order accuracy, by using the midpoint rule for surface integrals in combination with linear interpolation schemes. Moreover, the pressure field was computed on a staggered control volume, since large pressure gradients might appear due to the swirling motion. The steady problem was solved using a SIMPLE algorithm with under-relaxation, adjusted to increase the stability of the calculation by enhancing diagonal dominance of the coefficient matrix~\citep{Jasak96}. The discretization error was estimated through a \emph{Grid Convergence Index} study on three consecutively refined grids, as described in~\cite{Roache94}, as well as using the Richardson extrapolation, that assumes a monotonic convergence when grids are sufficiently fine~\citep{Richardson10}. The convergence index is defined as $\text{GCI}_{j+1,j}(\%)=3|(f_{j+1}-f_j)/[f_j(\alpha^n-1)]|\times 100$, where $f$ refers to any integral quantity or field value obtained using the coarse, $j+1$, or the fine, $j$, grids, $n = \log[(f_{j+2}-f_{j+1})/(f_{j+1}-f_j)]/log(\alpha)$ is the calculated convergence order, and $\alpha=(N_{j}/N_{j+1})^{1/2}$ the grid refinement ratio in each coordinate. In our problem, we computed $\text{GCI}_{j+1,j}$ for two field values at the near wake, namely $V(r=0.5,~z=1)$ and $W(r=0.5,~z=5)$, being the grid refinement ratio $\alpha\simeq\sqrt{2}$. Table \ref{GCI_BF} shows the values obtained for three consecutive grids, where it is seen how the grid convergence index is reduced as the number of nodes is increased, i.e., $\text{GCI}_{2,1}~<~\text{GCI}_{3,2}$. This result proves that the solution is converging towards the grid-independent one, and that further increments in $N$ will hardly improve the accuracy of the solution while increasing the computational time. Thus, grid $\#1$ was used in all the results reported herein to calculate the velocity and pressure fields for the base flow. On the other hand, as mentioned by~\cite{Ferziger02}, the estimated convergence order, $n$, depends on the grid resolution in the region where the magnitude used for the $\text{GCI}$ study is monitored. Consequently, since we focus on the near wake, the grids have been thoroughly refined in that region, providing values for the estimated order of the discretization schemes almost identical for both $V(r=0.5,z=1)$ and $W(r=0.5,z=5)$, namely $n_V\simeq2.067$ and $n_W\simeq2.000$, respectively, in agreement with the second-order discretization schemes set in the finite volume code.

Moreover, to validate the results obtained with the global linear stability analysis at high values of $\Omega$, direct numerical simulations were also performed. In this case, the full unsteady, three-dimensional Navier-Stokes equations~\eqref{NV1}-\eqref{NV2} were solved with the finite-volume software OpenFOAM\textsuperscript{\textregistered}. Specifically, spatial derivatives were discretized using a second-order linear interpolation for the diffusive term, whereas a total-variation-diminishing scheme with a van Leer limiter was selected for the convective term. Time integration was performed through a Crank Nicholson method blended with Euler integration, and the pressure-velocity coupling was tackled with a PISO algorithm. For a detailed description of the numerical methodology employed for the unsteady, three-dimensional simulations, as well as a thorough validation study, the reader is referred to~\cite{Jimenez13}.

\subsection{Global stability technique}
\begin{figure}
\begin{center}
\includegraphics[width=13cm]{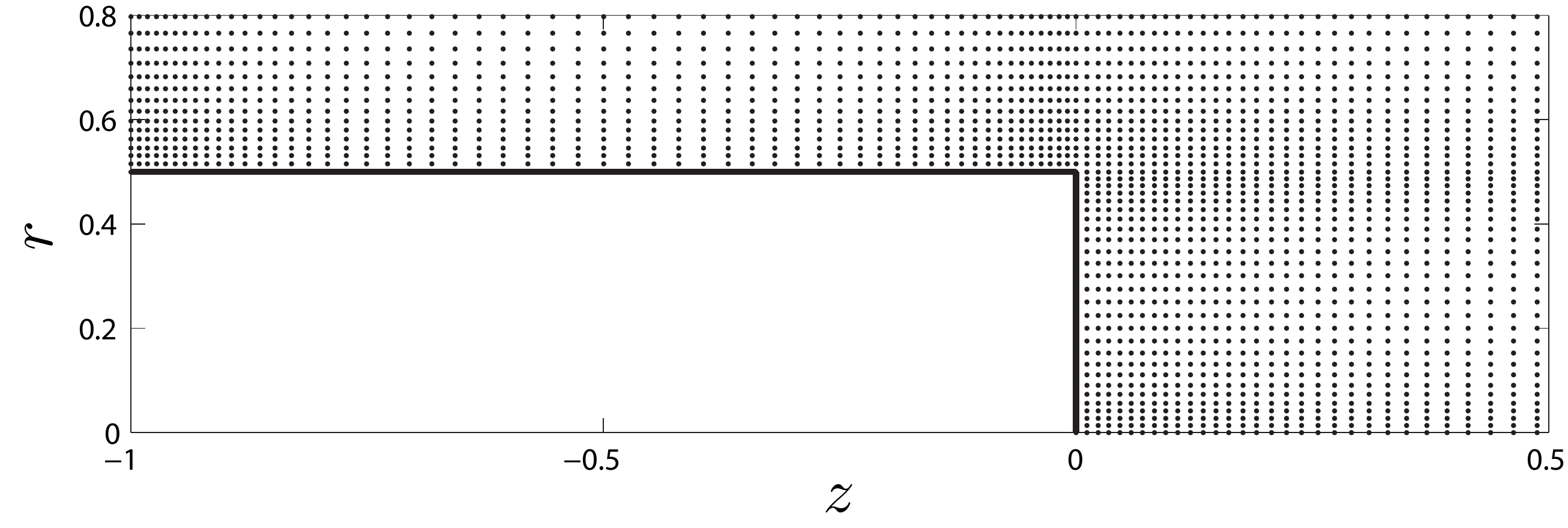}
\caption{Detail of the non-uniform grid used for the numerical solution of the global
stability problem.}
\label{grid}
\end{center}
\end{figure}

The global eigenmodes of the axisymmetric flow were obtained by discretizing equation~\eqref{eigenproblem} on a non-uniform grid of size $N$ (see detail shown in figure~\ref{grid}), whose rectangular domain extends from $z=-1$ to $z=50$, as depicted in figure~\ref{scheme}. To that end, the axial and radial derivatives appearing in the operator $\mathcal{A}$ (equation~\ref{matrixA}) were computed using sixth-order finite differences for non-uniform grids~\citep{Moin10,Sanmiguel11}. Since all the flow variables are evaluated on a collocated grid arrangement, filters derived from fourth-order compact schemes were used to avoid spurious oscillations of short length scale~\citep{Lele92}. A standard cubic interpolation algorithm was used to map the base flow variables on the non-uniform grid used in the stability analysis. The boundary conditions implemented in the global stability problem are,
\begin{gather}
\mathbf{\hat{u}}=(0,0,0),~~~\mathbf{x}~\in~\Sigma'_i \cup \Sigma_w \cup \Sigma_f ,\\
\hat{p}\,\mathbf{n}-\Rey^{-1}\mathbf{n}\cdot\nabla\mathbf{\hat{u}}=0,~~~\mathbf{x}~\in~\Sigma_o,
\end{gather}
treating the pressure implicitly by means of the incompressibility condition~\eqref{PF1} at each boundary. At the symmetry axis, $\Sigma_a$, the boundary conditions were obtained assuming that all physical quantities were smooth and bounded at $r = 0$ \citep[see][for details]{Khorrami89}.
\begin{table}
\centering
\begin{tabular}{c c c c c c c c}
Grid & $N$ & $n_r$ & $n_z$ & $\sigma_1$ & $\sigma_2$ & $\epsilon_{j,j+1}(\%)$ & $t_{\text{EVP}}(s)$\\ \\
 $1$ & $83000$ & $162$ & $520$ & $-0.02054+ \, 0.69332\, i$ & $-0.04280+\, 0.95797\, i$ & $0.12$ & $18900$\\
 $2$ & $63900$ & $142$ & $456$ & $-0.02052+ \, 0.69251\, i$ & $-0.04275+\, 0.95836\, i$ & $7.27$ & $9900$\\
 $3$ & $49200$ & $124$ & $400$ & $-0.02057+\, 0.69553\,  i$ & $-0.04586+\, 0.95955\, i$ & $4.60$ & $5800$\\
 $4$ & $38000$ & $109$ & $351$ & $-0.02146+\, 0.69713\, i$ & $-0.04797+\, 0.95949\, i$ & - & $3600$
\end{tabular}
\caption{Grid convergence study based on the two leading eigenvalues, $\sigma_1$ and $\sigma_2$
respectively, found for $\Omega=0.5$ and $\Rey=350$, for several meshes of size ratio
$N_{j}\simeq N_{j+1}\times 1.3$. Also shown are the maximum relative error,
$\epsilon_{j,j+1}(\%)=\left |\R(\sigma_2(j))-\R(\sigma_2(j+1))\right|/\R[\sigma_2(j)]\times 100$, and the
computational time needed to solve the EVP, $t_{\text{EVP}}(s)$.}
\label{GLSconvergence}
\end{table}

Once the operators stated in equation~\eqref{eigenproblem} are discretized and the boundary conditions imposed, the generalized EVP can be expressed as a set of linear algebraic equations,
\begin{equation}
\mathbf{A}\mathbf{\hat{q}}=\sigma\mathbf{B}\mathbf{\hat{q}},
\label{mateig}
\end{equation}
where $\mathbf{A}$ and $\mathbf{B}$ are $4N\times4N$ matrices which correspond to the discretized versions of the differential operators $\mathcal{A}$ and $\mathcal{B}$ given in equations~\eqref{matrixA} and~\eqref{matrixB}. Note that, the matrices $\mathbf{A}$ and $\mathbf{B}$ already incorporate the boundary conditions discussed above. Since an accurate numerical solution of the global stability problem requires values of $N\sim \mathcal{O}(10^4-10^5)$ (see table~\ref{GLSconvergence}), the QZ algorithm, aimed at recovering the full spectrum, is impractical. Hence, as in~\cite{Sanmiguel09}, the iterative Arnoldi method~\citep{Arnoldi51} was used to solve a standard EVP obtained from equation~\eqref{mateig} by means of a \emph{shift and invert} strategy~\citep{Theofilis11}, $(\mathbf{A}-\beta\mathbf{B})^{-1}\mathbf{B}\mathbf{\hat{q}}=1/(\sigma-\beta)\mathbf{\hat{q}}$, searching for the least stable part of the spectrum, i.e., the eigenvalues of largest magnitude in the vicinity of the shift parameter $\beta$. The eigenvalue computations were performed with standard routines available in MATLAB\textsuperscript{\textregistered}, that are based on the ARPACK library~\citep{Lehoucq98}. As pointed out by~\cite{Theofilis03}, the convergence provided by the Arnoldi method for the least stable eigenmodes depends on the value of $|\beta-\sigma_i|$, but mostly depends on the number of iterations of the Arnoldi method, $s$, i.e., the size of the Hessenberg matrix built from the Krylov basis. Thus, when $s$ is large enough, the influence of $|\beta-\sigma_i|$ vanishes.

To select the optimal value of $s$, a test was performed for the particular case of $\Rey=340$ and $\Omega=0.1$, and a shift parameter of $\beta=0.2+ 0  \,i$. Under these conditions, the numerical simulations revealed the periodic emission of vortical structures at an angular frequency $\sigma_i=2\pi St\simeq 0.0961$, in agreement with the fact that the leading eigenvalue provided by our global stability analysis, namely $\sigma_1=0.00630866+ 0.09210348 \, i$, is marginally unstable in this case. Moreover, the value of $\sigma_1$ was found to converge for $s\geq 75$ and, consequently, all the eigenvalue calculations reported herein were based on a conservative value of $s=100$, to obtain better convergence for the stable part of the spectrum.

A convergence study was also performed to check the accuracy of our results and to select the optimal grid size. Table~\ref{GLSconvergence} shows the values of the two leading eigenvalues found for $\Rey=350$ and $\Omega=0.5$, using four different grids, together with the associated computational time. The grid resolution was successively increased by a factor of $\sqrt{1.3}$ in each coordinate between consecutive grids, so that the number of nodes increased by a factor of approximately $1.3$. In this table $n_r$ and $n_z$ indicate the number of grid points in the radial and axial directions respectively. The column corresponding to $\epsilon_{j,j+1}(\%)$ shows the relative errors obtained between two consecutive grids for the growth rate of the next-to-leading eigenmode, $\R(\sigma_2)$, since this magnitude has the largest associated error for all grids, when compared to $\I(\sigma_{1,2})$ or $\R(\sigma_1)$. According to the results of table~\ref{GLSconvergence}, it can be deduced that the relative differences between the results obtained with grids \#1 and \#2 are very small as far as real and imaginary parts of the eigenvalues are concerned, with a maximum relative error of $0.12\%$. These small discrepancies, together with the fact that the computational time associated to grid \#$1$ is almost twice as large as that associated to grid \#2, motivated our selection of grid \#$2$ in all the calculations reported herein. As a minor comment, it should be noted that highly damped modes of the spectra were practically impossible to converge, in line with what occurs with other open flows, especially wakes~\citep[see for instance the investigation of the sphere wake by][where the authors study the influence of the location of the outer boundaries, and the mesh refinement, on the convergence of highly damped eigenmodes]{Natarajan93}.

\end{appendix}


\end{document}